\documentclass[11pt]{article}

\usepackage{color}

\usepackage{amsmath, amssymb,amsfonts,longtable,tabularx}
\usepackage{cite}
\usepackage{colortbl}
\usepackage[all]{xy}
\usepackage{epsfig}
\usepackage{subfigure}
\usepackage{graphics}
\usepackage{multirow}

\usepackage{setspace}
\usepackage{caption}
\usepackage{pifont}
\usepackage[table]{xcolor}

\definecolor{gray1}{rgb}{0.9,0.9,0.9}
\definecolor{gray2}{rgb}{0.8,0.8,0.8}

\usepackage[hmargin=2cm, vmargin=2cm]{geometry}

\date{}

\usepackage{cite}

\begin{document}

\begin{flushleft}
{\huge
\textbf{Small games and long memories promote cooperation}
}
\bigskip
\\
Alexander J. Stewart$^{1,2}$, 
Joshua B. Plotkin$^{1}$
\\
\bigskip
\bigskip
$^1$ Department of Biology, University of Pennsylvania, Philadelphia, PA 19104, USA
\\
$^2$ Current address: Department of Genetics, Environment and Evolution, University College London, London, UK
\\
\end{flushleft}

\noindent \textbf{Complex social behaviors lie at the heart of many of the challenges facing
evolutionary biology, sociology, economics, and beyond.  For evolutionary
biologists in particular the question is often how such behaviors can arise
\textit{de novo} in a simple evolving system. How can group behaviors such as
collective action, or decision making that accounts for memories of past
experience, emerge and persist? Evolutionary game theory provides a framework for
formalizing these questions and admitting them to rigorous study. Here we develop such a framework 
to study the evolution of sustained collective action in multi-player public-goods games, 
in which players have arbitrarily
long memories of prior rounds of play and can react to their
experience in an arbitrary way. To study this problem 
we construct a coordinate system for memory-$m$ strategies in 
iterated $n$-player games that permits us to characterize all 
the cooperative strategies that resist invasion by any mutant strategy, and thus
stabilize cooperative behavior. We show that while larger games inevitably make
cooperation harder to evolve, there nevertheless always exists a positive volume of strategies that
stabilize cooperation provided the population size is large enough. We also show
that, when games are small, longer-memory strategies make cooperation easier to
evolve, by increasing the number of ways to stabilize cooperation. Finally we
explore the co-evolution of behavior and memory capacity, and we find that longer-memory
strategies tend to evolve in small games, which in turn drives the evolution of
cooperation even when the benefits for cooperation are low.}
\\
\\
\\
Behavioral complexity is a pervasive feature of organisms that engage
in social interactions. Rather then making the same choices all the time -- always
cooperate, or never cooperate -- organisms behave differently depending on their
social environment or their past experience.  The need to understand
behavioral complexity is at the heart of many important challenges facing
evolutionary biology as well as the social sciences, or indeed any problem in which social 
interactions play a part. Cooperative social interactions in
particular play a central role in many of the major evolutionary transitions, from
the emergence of multi-cellular life to 
the development of human language \cite{Maynard-Smith:1995aa}.

Evolutionary biologists have been successful in pinpointing biological and
environmental factors that influence the emergence of cooperation in a population.
The demographic and spatial structure of populations in particular have emerged as fundamentally
important factors
\cite{Nowak:2006ys,Lieberman:2005aa,Hauert:2004aa,Rousset:2004aa,Nowak:2006ly,Komarova:2014aa,Gavrilets:2014aa}.
At the other end of the scale, the underlying mechanisms of cooperation -- such as 
the genetic architectures that encode social traits, or the ability of public goods to
diffuse in the environment -- also place constraints on how and to what
extent cooperation will evolve
\cite{Allen:2013aa,Menon:2015aa,Julou:2013aa,Cordero:2012vn,Axelrod:2006zr}. 

Despite extensive progress for simple interactions, an understanding of the evolution of
cooperation when social interactions occur repeatedly -- so that individuals can
update their behavior in the light of past experience -- and involve
multiple  participants simultaneously, remains elusive.  Some of the most promising
approaches for tackling this problem come from the study of iterated games
\cite{Nowak:1993vn,Nowak:2004fk,Imhof:2007uq,Sigmund:2010ve,Press:2012fk,Akin,Axelrod2,Von-Neumann:2007aa}.
In the language of game theory, behavioural updates in light of past experience
are modelled as a strategy in an iterated multi-player game among
heterogenous individuals. Even when we limit ourselves to a small set of
relatively simple strategies in such games, the resulting evolutionary dynamics 
are often surprising and counter-intuitive. As we begin to allow for
a wider array of ever more complex behaviors, results on the emergence of
cooperation are correspondingly harder to pin down.

In this paper we study 
evolving populations composed of individuals playing arbitrary strategies in iterated, multiplayer games.
We focus on the prospects for cooperation in public-goods games, and we
investigate how these prospects depend on the number of players that
simultaneously participate in the game, on the memory capacity of the players, and
on the total population size. We then study the co-evolution of players'
strategies alongside their capacity to remember prior interactions. We arrive at a
simple insight: when games involve few players, longer memory strategies tend to
evolve, which in turn increases the amount of cooperation that can 
persist. And so populations tend to progress from short memories and selfish
behavior to long memories and cooperation.


\section{Results}

We study the evolution of cooperation in iterated public-goods games, in which $n$
players repeatedly choose whether to cooperate by contributing a cost $C$ to a
public pool, producing a public benefit $B>C$. In each round of iterated play the total
benefit produced due to all player's contributions is divided equally between all
players. Thus, if $k$ players choose to cooperate in a given round, each player
receives a benefit $Bk/n$. We study finite populations of $N$ players engaging in
infinitely iterated $n$-player public-goods games, using strategies with memory
length $m$, meaning a player can remember how many times she and her opponents
cooperated across the preceding $m$ rounds (Figure 1). 

\begin{figure}[h!] \centering \includegraphics[scale=0.4]{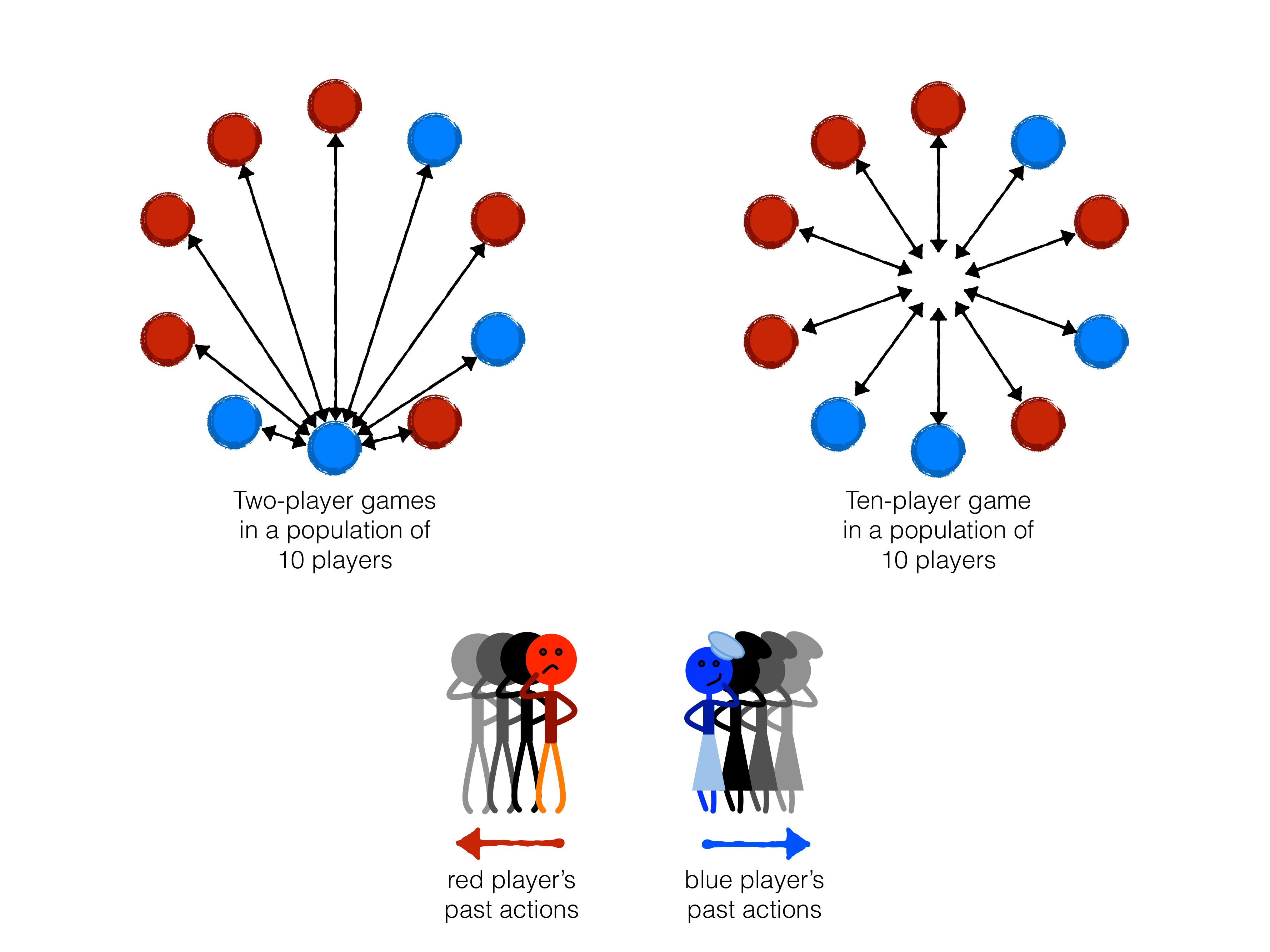}
\caption{Multiplayer games and memory. We study the evolution of behavior in
iterated $n$-player public-goods games in which players use strategies with memory
capacity $m$. We consider a replicating population of $N$ individuals who each
receive a payoff from engaging in an infinitely iterated game with all possible
subsets of $(n-1)$ opponents in the population.  Players then reproduce according to
a ``copying process'', in which a player $X$ copies another player's strategy $Y$
with a probability $f_{X\to
Y}=\frac{1}{1+\exp\left[\sigma\left(S^X-S^Y\right)\right]}$ where $S^X$ and $S^Y$
are the player's respective payoffs and $\sigma$ scales the strength of selection.
We consider the case of strong selection, such that a rare mutant who
is at a selective disadvantage is quickly lost from the population
\cite{Stewart:2014aa}. We investigate
the success of cooperative strategies as a function of game size and the length of
players' memories.  We determine the frequency of robust cooperative strategies,
which can resist invasion by any possible mutant.  (Top) Depending on the
size of the game $n$ relative to the population $N$, the dynamics of public-goods
games are different. In a two-player game, a series of pairwise interactions occur
in the population at each generation
(left). If the whole population plays the game each generation (right) all players interact
simultaneously.  (Bottom) Memory of past events results in strategies that update
behavior depending on the histories of both players' actions. This allows for more
complex strategies, such as those that punish rare defection or reward rare cooperation.}
\end{figure}

We focus on the evolution of sustained collective action, meaning the
evolution of strategies that, when used by each member of the population, produce
an equilibrium play with all players cooperating each round. This may be
thought of as the best possible social outcome of the game, because it produces 
the maximum total public good. We contrast the prospects for
sustained cooperation with the prospects for sustained \textit{inaction}, meaning
strategies that, when used by each member of the population, produce an
equilibrium play with all players defecting each round. This may be thought of
as the worst possible social outcome of the game, because it results in no public
good being produced at all.

To study the evolutionary prospects of collective action and inaction we determine
the ``volume of robust strategies'' that produce sustained cooperation or
defection in a repeated $n$-player game, in which players have memory $m$. The
game is played in a well-mixed population, composed of $N$ haploid individuals who
reproduce according to a ``copying process'' based on their payoffs (Figure 1)
\cite{Traulsen:2006zr}. The volume of robust strategies measures how much 
cooperation or defection will evolve across many generations
\cite{Stewart:2014aa}. More specifically, this volume is the probability that a
randomly drawn strategy that produces sustained cooperation (or defection) can
resist invasion by all other possible strategies that do not produce
sustained cooperation (or defection)
\cite{Stewart:2014aa,Stewart:2013fk,Stewart:2012ys,Akin,g6030231}.  As we have
shown previously \cite{Stewart:2014aa}, the volumes of robust strategies determine
the evolutionary dynamics of cooperation and defection in iterated games. We
confirm the utility of this approach by comparing our analytical predictions to
Monte Carlo simulations, studying the effects of population size, game size, and
memory capacity on the evolution of cooperation.

We begin our analysis by describing a coordinate system under which the volume of
robust strategies can be determined analytically, for games of size $n$, played in
populations of size $N$, in which strategies have memory length $m$. 
We use this coordinate system to completely characterize all evolutionary robust
cooperating (and defecting) strategies, which cannot be invaded by any
non-cooperating (or non-defecting) mutants, in the iterated $n$-player
public-goods game. We apply these results to make specific predictions for the
effects of game size and of memory capacity on the evolution of collective action
through sustained cooperation. Finally we explore the
consequences of these predictions for the co-evolution of cooperation and memory
capacity itself.

\subsection{Beyond two-player games and memory-1 strategies}

Recently, Press and Dyson introduced so-called zero determinant (ZD) strategies in
iterated two-player games \cite{Press:2012fk}.  ZD strategies are of interest
because, when a player unilaterally adopts such a strategy she enforces a
linear relationship between her longterm payoff and that of her opponent, and
thereby gains some measure of control over the outcome of the game
\cite{Hilbe:2013aa,Hilbe:2013uq,Hilbe:2014aa,Hilbe:2015aa,Hilbe:2015ab}. Several
authors have worked to extend the framework of Press and Dyson to multi-player
games \cite{Hilbe:2014aa,Pan:2015aa} and have characterized multi-player ZD
strategies, revealing a number of interesting properties.

Other research has expanded the framework of Press and Dyson to study all possible
memory-1 strategies for infinitely repeated, two-player games
\cite{Stewart:2014aa,Stewart:2013fk,Stewart:2012ys,Akin,g6030231}.  This work
involves developing a coordinate system for the space of all memory-1 strategies
\cite{Akin} that allows us to describe a straightforward (although not necessarily
linear) relationship between the two players' longterm payoffs. This relationship
between players' longterm payoffs, in turn, has enabled us to fully characterize
all memory-1 Nash equilibria and all evolutionary robust strategies for infinitely
repeated two-player games, played in a replicating population of $N$ individuals
\cite{Stewart:2014aa,Stewart:2013fk,Akin,g6030231}.

Here we generalize this body of work by developing a coordinate system for the
space of memory-$m$ strategies in multi-player games of size $n$, such that all
$n$ players' longterm payoffs are related in a straightforward (although not
necessarily linear) way.  One essential trick that enables us to achieve this goal
is to construct a mapping between memory-$m$ strategies in an $n$-player game and
memory-1 strategies in an associated $n\times m$-player game. We then construct a
coordinate system for the space of memory-1 strategies in multi-player games that
allows us to easily characterize the cooperating and the defecting strategies that
resist invasion. We apply these techniques to the case of iterated $n$-player
public-goods games and we precisely characterize all evolutionary robust
memory-$m$ strategies -- i.e.~those strategies that, when resident in a finite
population of $N$ players, can resist selective invasion by all other possible
strategies -- thereby elucidating the prospects for the evolution of cooperation
in a very general setting.

\subsection{A coordinate system for long-memory strategies in multi-player games}

Our goal is to study the effects of game size and memory on the frequency and
nature of collective action in public-goods games. Allowing for
long-memory strategies and games with more than two players greatly expands
the potential for behavioral complexity, because  players are able to react to the
behaviors of multiple opponents across multiple prior interactions. And so merely
determining the payoffs received by players in such an iterated public-goods game
can pose a significant challenge. In order to tackle this problem we develop a
coordinate system for parameterizing strategies, in which the outcome of a game
between multiple players using long-memory strategies can nonetheless be easily
understood.

A player using a memory-$m$ strategy chooses her play in each round of an
iterated game
in a way that depends on the history of plays by all $n$ players across the
preceding $m$ rounds. In general such a strategy consists of $2^{n\times m}$
probabilities for cooperation in the present round. We write the probability for
cooperation of a focal player in its most general form as
$p_{\boldsymbol{\sigma}_0,\boldsymbol{\sigma}_1,\boldsymbol{\sigma}_2,\ .\ .\ .\
,\boldsymbol{\sigma}_{n-1}}\in[0,1]$ where $\boldsymbol{\sigma}_i$ denotes the
history of plays for player $i$. Each $\boldsymbol{\sigma}_i$ corresponds to an
ordered sequence of $m$ plays for player $i$, with each entry taking the value
$\textbf{c}$ (cooperate) or $\textbf{d}$ (defect). The $2^{n\times m}$
probabilities for cooperation form a basis for $\mathbb{R}^{2^{n\times m}}$ and
constitute a system of coordinates for the space of memory-$m$ strategies in
$n$-player games. In the supporting information we describe in detail how to
construct of an alternate coordinate system of $2^{n\times m}$ vectors that
also form a basis for $\mathbb{R}^{2^{n\times m}}$, and which greatly simplifies
the analysis of long-term payoffs in iterated games. Below we describe this alternative
coordinate system for the specific case of iterated public-goods games, which are the focus of this study.  
\\ 
\\ 
\textbf{(i) Mapping memory-$m$ to memory-1:} In order to simplify our analysis of long-memory strategies we will
conceive of a focal player using a memory-$m$ strategy in an $n$-player game as a
player who instead uses a memory-1 strategy in an associated $n \times m$-player
game. That is, we will think of an $n$-player game in which a focal player uses a
memory-$m$ strategy in terms of an equivalent $n\times m$-player game, which is
composed of $n$ ``real'' players along with $m-1$ ``shadow'' players associated
with each real player. The shadow players  play the same way that their associated
real player did $t$ rounds previously, for $2 \leq t \leq m$. The focal player's
memory-$m$ strategy is thus identical to a memory-1 strategy in the $n\times m$
player game, where the corresponding memory-1 strategy responds to a large set of
``shadow" players whose actions in the immediately previous round simply encode the
actions taken by the $n$ real players in the preceding $m$ rounds. This trick
allows us to reduce the problem of studying long-memory strategies to
the problem of studying memory-1 strategies, albeit with a larger number
of players in the game.

All that is required is to construct strategies for the shadow players so that the
state of the system across the preceding $m$ rounds is correctly recreated at each
round of the associated $n\times m$-player game. This construction is straight forward. If
the focal player played $c$ in the last round, then we stipulate that her first
shadow player will play $c$ in the next round (i.e.~it will copy her last move).
Similarly her second shadow player will copy the last move of her first shadow
player, and so on, up to her $(m-1)$st shadow player. The same goes for the shadow
players of each of her $n-1$ opponents. In this way, all the plays of the last $m$
rounds in the $n$-player game are encoded at each round in the associated $n
\times m$-player game.

Having transformed an arbitrary memory-$m$ strategy in an $n$-player into an
associated memory-1 strategy in an $n\times m$-player game, we now describe a
coordinate system for memory-1 strategies that allows us to derive a simple
relation among the equilibrium payoffs to all players. We define this coordinate
system for arbitrary games in the supporting information (section 3), and for the
case of public-goods games below.
\\
\\
\textbf{(ii) Parameterizing strategies in public-goods games:}
Under a public-goods game, a player who cooperates along with $k$ of her opponents
receives a net payoff $Bk/n-C$, whereas a player who defects while $k$ of her
opponents cooperate receives a net payoff $Bk/n$. That is, the payoff received
depends on whether or not the focal player cooperated and on the number of her
opponents that cooperated, but it does not depend on the identity of her
cooperating opponents. Likewise, if a player has memory of the preceding $m$
rounds of an iterated public-goods game, then her payoff across those rounds
depends on the total number of times she cooperated and the total number of times
her opponents cooperated, but it does not depend on the order in which different
players cooperated nor on the identity of her cooperating opponents. Therefore,
rather than studying the full space of $2^{n\times m}$ probabilities for
cooperation, we can limit our analysis for iterated public-goods games to
strategies that keep track of the total number of times a focal player cooperated,
and the number of times her opponents cooperated, within her memory capacity. A focal
player's strategy can thus be expressed as $((n-1)m+1)\times(m+1)$ probabilities
for cooperation each round, $p^{l_o,l_p}$, where $l_o$ denotes the total of number
of times the player's opponents cooperated in the preceding $m$ rounds (which
number can vary between 0 and $(n-1)m$) and $l_p$ denotes the total number of
times the player herself cooperated in the preceding $m$ rounds (which can vary
between 0 and $m$).

\begin{figure}[h!] \centering \includegraphics[scale=0.4]{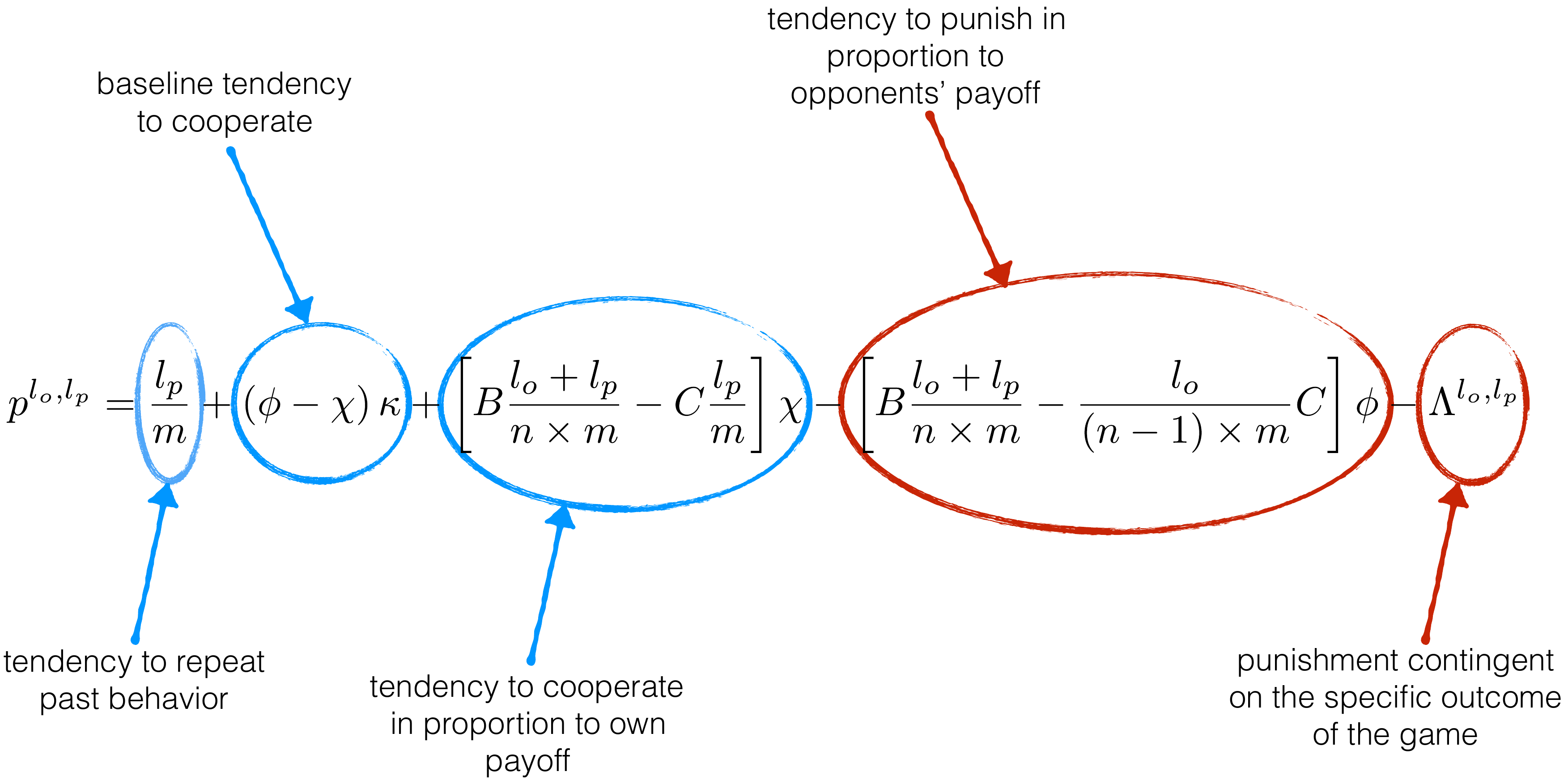}
\caption{A co-ordinate system for describing strategies in public-goods games.  We consider the
space of strategies of the form $p^{l_o,l_p}$, such that players cooperate with a
probability that depends on the number of times $l_o$ her opponents have cooperated and
the number of times $l_p$ she has cooperated within her memory. 
We define the strategy of a focal player by coordinates $\{\chi,\phi,\kappa,
\Lambda^{0,0},\ldots\Lambda^{(n-1)\times m,m}\}$ as shown in the figure.
The components of this coordinate system have an intuitive interpretation: the probability that a player
cooperates depends on (1) her past tendency to cooperate, (2) a baseline tendency
to cooperate ($\kappa$), (3) a tendency to cooperate in proportion to her own payoff ($\chi$), (4) a
tendency to punish (i.e.~defect) in proportion to her opponents' payoffs ($\phi$) and (5) a
contingent punishment that depends on the specific outcome of the game over the
prior $m$ rounds ($\Lambda^{(n-1)\times m,m}$).} \end{figure}

Although the probabilities $p^{l_o,l_p}$ are perhaps the most natural coordinates
for describing a memory-$m$ strategy in an iterated $n$-player public-goods game,
we have developed an alternative coordinate system, defined in Figure 2, that
simplifies the analysis of equilibrium payoffs and the evolutionary robustness of
strategies. The alternative system of $((n-1)m+1)\times(m+1)$
coordinates for a given player's strategy is described by parameters
$\{\chi,\phi,\kappa,\Lambda^{0,0},\ldots,\Lambda^{(n-1)\times m,m}\}$
defined in Figure 2. We impose the boundary conditions 
${\Lambda}^{0,0}={\Lambda}^{(n-1)\times m,m}=0$ along with one other linear
relationship on the $\Lambda$ terms (see supporting information).
Qualitatively, this coordinate system describes the
probability of cooperation in a given round, $p^{l_o,l_p}$, in terms of a weighted
sum of five components: (1) The tendency to repeat past behavior; (2) The baseline
tendency to cooperate ($\kappa$); (3) The tendency to cooperate in proportion to
the payoff received by the focal player ($\chi$); (4) The tendency to punish (i.e.~defect) in
proportion to the payoffs received by her opponents ($\phi$) and (5) The tendency
to punish in response to the specific outcome of the previous rounds
($\Lambda^{l_o,l_p}$). 

The advantage of using this coordinate system is that it provides a simple
relationship between the long-term payoff to a focal player $0$, $S^0$, and the
the long-term payoffs $S^i$ of each of her opponents $i$ in an iterated $n$-player
public-goods game:
\begin{equation}
\phi\sum_{i=1}^{n-1}\frac{S^i}{n-1}-\chi
S^{0}-\kappa(\phi-\chi)+\sum_{l'_o=0}^{(n-1)\times m}\sum_{l'_p=0}^{m}\hat{\Lambda}^{l'_o,l'_p}w^{l'_o,l'_p}=0.
\end{equation}
Here the term $w^{l'_o,l'_p}$ denotes the equilibrium rate at which the invading player
cooperates $l'_p$ times and his opponents cooperate $l'_o$ times over the preceding
$m$ rounds, and $\hat{\Lambda}^{l'_o,l'_p}$ denotes the contingent punishment of the 
focal strategy from the point of view of a mutant (see supporting information for a derivation of 1).
$\hat{\Lambda}^{l'_o,l'_p}$ is related in a
simple way to the terms $\Lambda^{l_o,l_p}$, 
so that increasing $\Lambda^{l_o,l_p}$ increases $\hat{\Lambda}^{l'_o,l'_p}$ (see supporting information).

\subsection{The effects of game size on robust cooperation} The relationship among
payoffs summarized in 1 provides extensive insight into the outcome of
iterated public-goods games. Of particular interest are the prospects for 
cooperation as the game size $n$ and population size $N$ grow.
Public-goods games are well known examples of the collective action problem, in
which increasing the number of players in a game worsens the prospects for
cooperation \cite{Ostrom:1990aa,Gavrilets:2015aa}. Larger populations, on the
other hand, tend to make it easier to evolve robust cooperation, at least for two-player
games \cite{Stewart:2013fk}.  We will use 1 to explore the tradeoff between
game size and population size, and the nature of robust cooperative behaviors that
can evolve in multi-player games.
  
1 allows us to characterize the ability of a cooperative strategy
to resist invasion by any other strategy in a population of size $N$\cite{Stewart:2014aa,Stewart:2013fk,Akin,g6030231}. We define a cooperative
strategy as one which, when played by every member of a population, assures that
all players cooperate at equilibrium and thus receive the payoff for mutual
cooperation, $B-C$. This implies the necessary condition $p^{(n-1)\times m,m}=1$,
so that if all players cooperated in the preceding $m$ rounds, a player using a
cooperative strategy is guaranteed to cooperate in the next round. We call such
strategies ``cooperators'' meaning that they produce sustained cooperation when resident in a
population. In the alternate coordinate system developed above a necessary condition for
sustained cooperation is $\kappa=B-C$.  

Conversely, we also consider strategies
that lead to collective \textit{inaction}, meaning sustained
defection. Such strategies must have $p^{0,0}=0$, which implies a necessary
condition $\kappa=0$ in the alternate coordinate system. We call strategies
satisfying this condition ``defectors'' meaning that they produce sustained
defection when resident in a population. 

A rare mutant $i$ can invade a population of size $N$ in which a cooperative strategy is
resident only if he receives a payoff $S^i$ that exceeds the payoff received by the resident cooperator. 
By considering bounds on the
payoffs received by players (see supporting information)
we have derived necessary and
sufficient conditions for a cooperative strategy $\{\chi,\phi,\kappa,
\hat{\Lambda}^{0,1},\ldots\hat{\Lambda}^{(n-1)\times m,m-1}\}$ 
to resist selective invasion by any mutant strategy -- that is, for a cooperative
strategy to be evolutionary robust:
\footnotesize 
\begin{align*}
\nonumber  &\mathcal{C}^{n,m}_{s}=\Bigg\{(\chi,\phi,\kappa,
\hat{\Lambda}^{0,0},\ldots\hat{\Lambda}^{(n-1)\times m,m})\bigg|\kappa=B-C,\\
\nonumber &\frac{N-n}{N-1}\sum_{l_o=0}^{(n-1)\times m}\sum_{l_p=0}^{m}\hat{\Lambda}^{l_o,l_p}w^{l_o,l_p}\geq \\
\nonumber & C\left(\phi\frac{N(n-2)+1}{(N-1)(n-1)}-\chi\right)
\sum_{l_o=0}^{(n-1)\times m}\sum_{l_p=0}^{m}\frac{l_o+l_p}{(n-1)\times m}w^{l_o,l_p},\\
\nonumber &\frac{N-n}{N-1}\sum_{l_o=0}^{(n-1)\times m}\sum_{l_p=0}^{m}\hat{\Lambda}^{l_o,l_p}w^{l_o,l_p}\geq \\
\nonumber &(B-C)\left(\phi\frac{N(n-2)+1}{(N-1)(n-1)}-\chi\right)
\sum_{l_o=0}^{(n-1)\times m}\sum_{l_p=0}^{m}\frac{n\times m-l_o-l_p}{(n-1)\times m}w^{l_o,l_p}\Bigg\}.
\end{align*}
\normalsize
\begin{equation}
\end{equation}

2 allows us to make a number of observations about the prospects and nature of
robust cooperation. First, all other things being equal, larger values of
$\hat{\Lambda}^{l_o,l_p}$, which correspond to stronger contingent punishment, in which
players successfully punish rare defection, make it easier for a strategy to
satisfy the
requirements for robust cooperation.  Second, positive values of $\chi$,
corresponding to more generous strategies \cite{Stewart:2013fk}, in which players tend to share the benefits
of mutual cooperation, also make it easier for a strategy to satisfy the requirements for
robust cooperation.  Thus, complex strategies that punish rare defection and are
generous to other players tend to produce robust cooperative behavior in an
evolving population.

2 also shows that larger values of $n$, corresponding to games with more
players, tend to make for smaller volumes of robust cooperative strategies.  This
can be see on the left-hand side of the inequality in 2, where increasing $n$
attenuates the impact of contingent punishment on robustness. Likewise, this can
also been seen on the right-hand side of the iniquality in 2, where increasing
$n$ attenuates the impact of generosity on robustness. 

The effects of game size on the prospects for cooperation can be illustrated by
considering two extreme cases.  When the entire population takes part in a single
multi-player game, so that $n=N$, then 2 implies that strategies can be
robust only if $\chi\geq\phi$. However, in order to produce a viable strategy 
$\chi\leq \phi$ is required (Fig.~2); and so the only possible way to ensure robust
cooperation in this extreme case is to have $\chi=\phi$.  The condition
$\chi=\phi$ gives a tit-for-tat-like strategy, and it results in unstable
cooperative behavior in the presence of noise \cite{Stewart:2014aa}. And so, in
the limit of games as large as the entire population size the prospects for
evolutionary robust cooperation are slim.  However, in the contrasting case in
which the population size is much larger than the size of the game being
played, that is $N\gg n\gg1$, then 2 shows that a positive
volume of robust cooperative strategies always exists, given sufficient contingent
punishment $\Lambda^{l_o,l_p}$, even in very large games.

Understanding the expected rate of cooperation in multi-player games requires that
we compare the volume of robust cooperative strategies to the volume of robust
defecting strategies. A rare mutant $i$ can invade a population in which a
defecting strategy is resident only if he receives a payoff $S^i$ that exceeds the payoff received by the resident defector.  The
resulting necessary and sufficient conditions for the robustness of defecting
strategies are then:
\footnotesize
\begin{align*}
\nonumber  &\mathcal{D}^{n,m}_{s}=\Bigg\{(\chi,\phi,\kappa,
\hat{\Lambda}^{0,0},\ldots\hat{\Lambda}^{(n-1)\times m,m})\bigg|\kappa=0,\\
\nonumber &\frac{N-n}{N-1}\sum_{l_o=0}^{(n-1)\times m}\sum_{l_p=0}^{m}\hat{\Lambda}^{l_o,l_p}w^{l_o,l_p}\geq\\
\nonumber&-(B-C)\left(\phi\frac{N(n-2)+1}{(N-1)(n-1)}-\chi \right)\sum_{l_o=0}^{(n-1)\times m}\sum_{l_p=0}^{m}\frac{l_o+l_p}{(n-1)\times m}w^{l_o,l_p},\\
\nonumber &\frac{N-n}{N-1}\sum_{l_o=0}^{(n-1)\times m}\sum_{l_p=0}^{m}\hat{\Lambda}^{l_o,l_p}w^{l_o,l_p}\geq\\
\nonumber &-C\left(\phi\frac{N(n-2)+1}{(N-1)(n-1)}-\chi\right)\sum_{l_o=0}^{(n-1)\times m}\sum_{l_p=0}^{m}\frac{n\times m-l_o-l_p}{(n-1)\times m}w^{l_o,l_p}\Bigg\}.\\
\end{align*}
\normalsize
\begin{equation}
\end{equation}
Once again, we see from 3 that larger values of $\hat{\Lambda}^{l_o,l_p}$, resulting
in stronger contingent punishment of rare cooperators in a population of
defectors, makes it easier for a defecting strategy to be robust. However, in contrast
to the case for cooperators, \textit{de}creasing $\chi$, which for defectors
corresponds to more extortionate behavior, such that players try to increase their
own payoff at their opponents' expense \cite{Press:2012fk}, makes a defecting
strategy more likely to satisfy the requirements for robustness. Finally, while
larger values of $n$ attenuate the effect of contingent punishment on robustness,
they also make more extortionate strategies more robust; and the latter effect is
always stronger, so that larger games permit a greater volume of defecting
strategies.  In the extreme case of $n=N$ all defecting strategies are robust.
Overall, 3 implies that increasing game size $n$ tends to increase the volume
of robust defectors, in contrast to its effect on robust cooperators. 

We confirmed our predictions for the effects of game size on the volume of
robust cooperators and defectors by analytical calculation of robust volumes, from
Eqs.~2-3, and by comparison to direct simulation for the invasibility of cooperators
and defectors against a large range of mutant invaders (Figure 3a).  As game size
increases the volume of robust cooperators decreases relative to the volume of
robust defectors, making cooperation harder to evolve. 

There is a simple intuition for why larger games make cooperation less robust 
and defection more robust: In public-goods games with more players,
the marginal change in payoff to a player who switches from cooperation to
defection is $C-B/n$, 
and so the incentive to defect grows as the size of
the game grows.  This of course is the group size paradox, and it is a well known
phenomenon for any collective action problem \cite{Ostrom:1990aa}. In the limiting case $n=N$ the only hope
for robust cooperation is tit-for-tat-like strategies, that are capable of both
sustained cooperation and sustained defection, depending on their opponent's
behavior. 

In general, both cooperators and defectors have positive volumes of robust strategies, provided $n<N$.  As such, both cooperation and defection can
evolve. Although these robust strategies cannot be selectively invaded by any other strategy when resident in a population, they can be neutrally
replaced by a non-robust strategy of the same type, which can in turn be selectively invaded.  As a result, there is a constant turnover between cooperation
and defection over the course of evolution, with the relative time spent at cooperation versus defection determined by their relative volumes of robust strategies
\cite{Stewart:2014aa,g6030231}.

Our results show that the problem of collective action is
alleviated by sufficiently large population sizes. That is, for an arbitrarily large
game size $n$ we can always find yet larger population sizes $N$ such that robust 
cooperative strategies are guaranteed to exist. 
Moreover, increasing the population size $N$ leads to increasing volumes of robust
cooperative strategies and decreasing volumes of robust defecting strategies
(Figure S1).  


\begin{figure}[h!] \centering \includegraphics[scale=0.4]{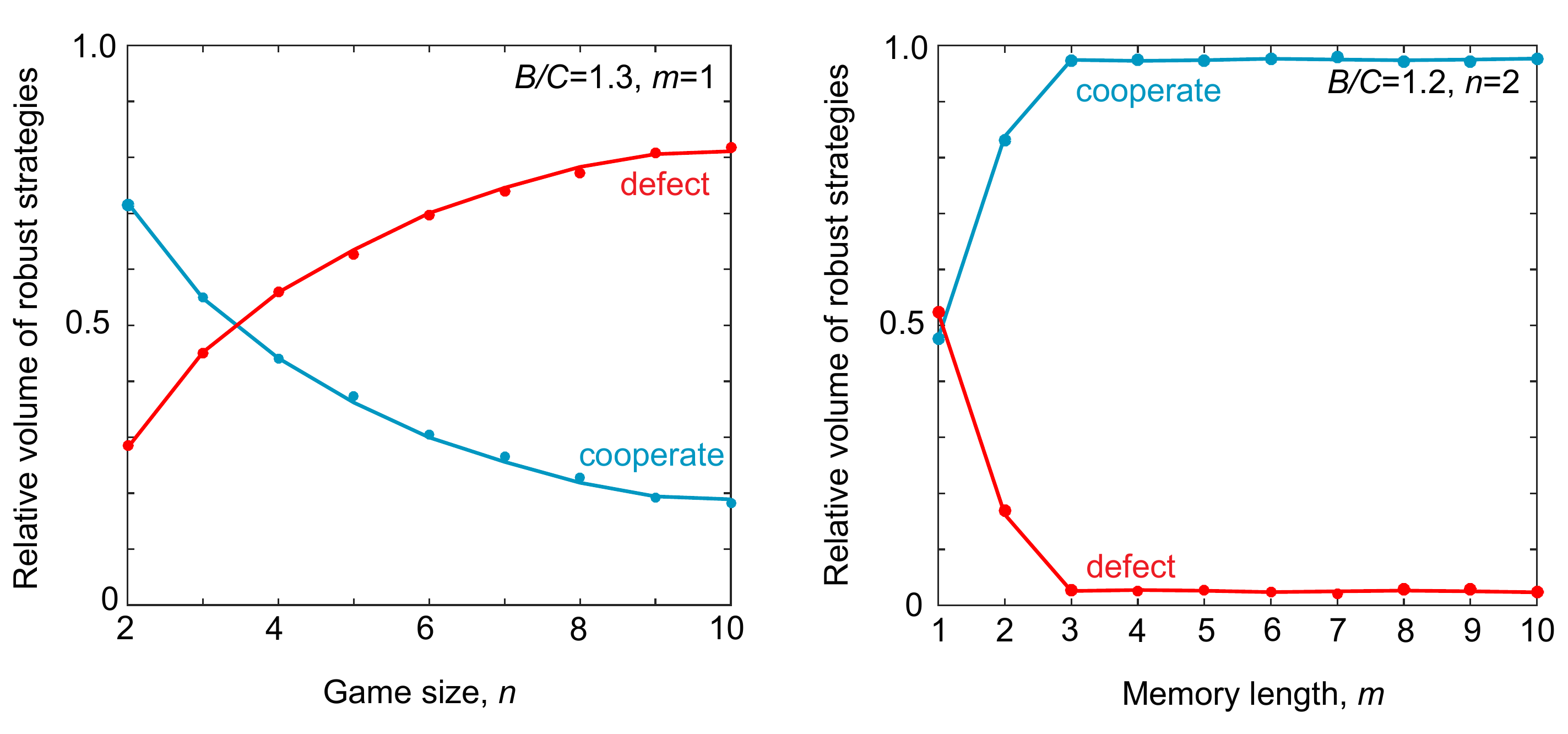}
\caption{The impact of game size and memory capacity on cooperation. We 
calculated the relative volumes of robust cooperation -- that is, the absolute
volume of robust cooperative strategies divided by the total volume of robust
cooperators and defectors -- and compared this to the relative volume of
defectors (solid lines) using Eqs.~2-3. We also
verified these analytic results by randomly drawing $10^6$ strategies and
determining their
success at resisting invasion from $10^5$ random mutants (points). We calculated player's
payoffs by simulating $2\times 10^3$ rounds of a public-goods game. We then
plotted the relative volumes of robust cooperators and robust defectors as a
function of game size $n$ (with fixed memory $m=1$, left) and as a function of memory
capacity $m$ (with fixed game size $n=2$, right). Increasing game size increases
the relative volume of robust defection; while increasing memory length
increases the relative volume of robust cooperation.
In all calculations and simulations we used cost $C=1$ and benefit $B$
as indicated in the figure.} \end{figure}

\subsection{The effects of memory on robust cooperation}

We have not yet said anything about the impact of memory capacity on the prospects
for cooperation. Indeed, the robustness conditions Eqs.~2-3 do not depend
explicitly on memory length $m$, as they do on game size $n$ and on population
size $N$.  However, memory does have an important impact on the efficacy of
contingent punishment, $\hat{\Lambda}$, on the left-hand sides of the inequalities
in 2 and Eq.~3.  Figure 3 illustrates the impact of increasing memory $m$ on
the volume of robust cooperative and robust defecting strategies. Here we see the
opposite pattern to the effect for game size: as memory increases, there is a larger
volume of robust cooperation relative to robust defection.

We can develop an intuitive understanding for the effect of memory on sustained
cooperation by considering its role in producing effective punishment. A longer
memory enables a player to punish opponents who seek to gain an advantage through
rare deviations from the social norm: that is, rare defectors in a population of
cooperators or rare cooperators in a population of defectors. However, using a
long memory to punish rare defectors is a more effective way to enforce
cooperation than punishing rare cooperators is to enforce defection (since in the
latter case the default behavior is to defect anyway, and so increasing the amount
of ``punishment'' has little overall effect on payoff). And so as memory
increases, cooperators become more robust relative to defectors, as 2-3 and
Figure 3 show.

The change in the efficacy of punishment for rare deviants from the social norm as
memory capacity increases is illustrated in Figure S2, where we calculate the
average $\hat{\Lambda}^{l_o,l_p}$ for randomly-drawn cooperative or defecting
strategies.  We see that as memory capacity increases, a randomly drawn cooperator tends to
engage in more effective punishment (larger values of $\hat{\Lambda}^{l_o,l_o}$)
whereas a randomly drawn defector tends to engage in less effective punishment
(smaller values of $\hat{\Lambda}^{l_o,l_o}$). This trend explains why 
increasing memory capacity increases the volume of robust cooperators relative
to defectors.

\subsection{Evolution of memory}

Our results on the relationship between memory capacity and the robustness of
cooperation raise a number of interesting questions. In particular, memory of the
type we have considered does not seem to convey a direct advantage to
cooperation (or defection), because a robust cooperative (or defecting) strategy
is robust against \textit{all possible invaders}, regardless of their memory
capacity. However increased memory can nonetheless make robust cooperation easier
to evolve, because it allows for more effective contingent punishment.  This tends
to have a stronger impact when games are small because, as described  in Eqs.
2-3,  the impact of contingent punishment on robustness is attenuated by a factor
$N-n$, and thus the effect of longer memory on the contributions 
of $\hat{\Lambda}$ terms to robust cooperation is smaller in larger games. And so, 
at least when the number of players is relatively small, we might expect long memories
to facilitate the evolution of cooperation in populations.

What our analysis has not yet addressed is whether memory capacity itself can
adapt, and what its co-evolution with strategies in a population will imply for
the longterm prospects of cooperation.  To address this question we undertook evolutionary
simulations, allowing heritable mutations both to a player's strategy and also to her
memory capacity. These simulations, illustrated in Figure 4, confirm that (i) 
longer memories do indeed evolve and (ii) this leads to an increase in the
amount of cooperation in a population (Figure 4). In a two-player game,
if memory has no cost, memory tends to increase over time, which in turn drives an
increase in the frequency of cooperators and a decline in defectors.  This is
accompanied by a large overall increase in the population mean fitness.  By
contrast, when the game size is large, $n=N$, there is little evolutionary change
in memory capacity and defection continues to be more frequent than cooperation
even as strategies and memory co-evolve.  When memory comes at a cost (Figure S3),
an intermediate level of memory evolves for small $n$, and there is a
correspondingly weaker increase in the degree of cooperation.

How are we to understand why memory evolves at all in these co-evolutionary
simulations? The change in memory capacity is puzzling, at first glance, because a
longer memory conveys no direct advantage against a resident robust strategy -- since
robustness implies uninvadability by any opponent, regardless of the opponent's
memory capacity.  The key to understanding this co-evolutionary pattern is to note
that longer memories are, on average, better at \textit{invading} non-robust
strategies, due to their greater capacity for contingent punishment (Figure S3).
Thus, when games are sufficiently small, the neutral drift that leads to turnover
between cooperation and defection \cite{Stewart:2014aa,g6030231} also provides
opportunity for longer-memory strategies to invade and fix.

\begin{figure}[h!] \centering \includegraphics[scale=0.33]{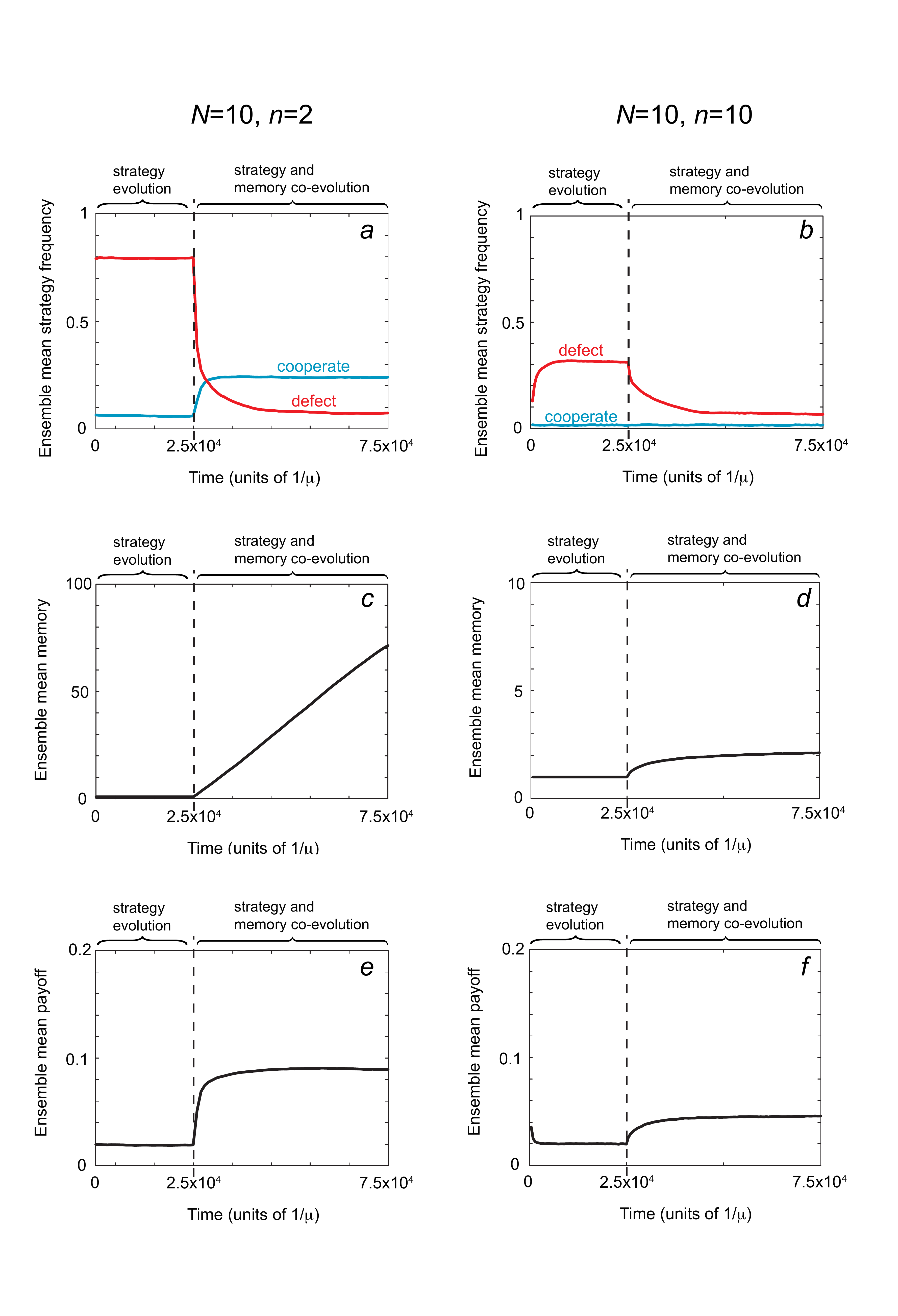}
\caption{Co-evolution of strategies and memory capacity.  We simulated populations
playing the iterated $n$-player public-goods game, proposing mutant strategies
until reaching equilibrium, and then also proposing mutations to a player's
memory capacity $m$, each at rate $\mu/10$. In these simulations all players
initially have memory $m=1$, with payoff parameters $C=1$ and $B=1.2$. Mutations
to strategies were drawn uniformly from the full space of memory-$m$ strategies.
Mutations perturbing the memory $m$ caused it to increase or decrease by $1$, with
a lower bound of $m=1$. Evolution was modeled according to a copying process under
weak mutation \cite{Stewart:2014aa} in a population of size $N=10$ individuals. (a) When
the game size is small, $n=2$, defecting strategies are initially dominant in the
population, but they are quickly replaced by cooperators as memory capacity
evolves to higher values.  (b) When game size is large, $n=N=10$, defecting
strategies initially dominate the population and they remain dominant as
memory evolves. In both (a) and (b) the overall frequency of cooperators and
defectors decline as the dimension of strategy space increases, in line with the
decline in the overall volume of robust strategies (Figure S4). (c) When the game
size is small memory
evolves rapidly to larger values, reflecting the greater success of longer-memory
strategies at invading (Figure S3), and driving the increase in
cooperative as compared to defecting strategies.  (d) When the game size is large
memory does not evolve to large
values, reaching only $m=2$ across 50,000 generations, and reflecting the decline in
long-memory strategies' success as invaders in larger games.  (e) As cooperation
increases so does the average payoff of the population, by a factor of 5-10 fold.
(f) The lack of increase in cooperation results in a much more modest (although
still appreciable) increase in average payoff for the population as defectors
become less frequent.  } \end{figure}

\clearpage


\section{Discussion}

We have constructed a coordinate system that enables us to completely characterize
the evolutionary robustness of arbitrary strategies in iterated multi-player public-goods
games. This allows us to quantify the contrasting impacts of the number of players
who engage in a game, and the memory capacity of those players, on the evolution
of cooperative behavior and collective action. In particular we have shown that
while increasing the number of players in a game makes both cooperation and longer
memories harder to evolve, in small games, memory capacity tends to increase over
time and drives the evolution of cooperative behavior.


To understand the evolution of social behavior it is not sufficient to simply
determine whether particular types of strategies exist or not. Indeed, for
repeated games, strategies that enforce any given social norm are guaranteed to
exist by the famous Folk Theorems \cite{Fund1}. The more incisive question, from
an evolutionary viewpoint, is how often strategies of different types arise via
random mutation, how often they reach fixation, and how long they remain fixed in
the face of mutant invaders and other evolutionary forces such as neutral genetic
drift. To address these questions we have analyzed the evolutionary robustness of
strategies that result in sustained cooperation. We have shown that a strategy is
more likely to be evolutionary robust if it can successfully punish defectors. We
have shown that players with longer memories have access to a greater volume of
such evolutionary robust strategies, and that, as a result, over the course of
evolution populations that evolve longer memories are more likely to evolve
cooperative behaviors. Memory of the type we have considered does not result in
better strategies per se, but in a greater quantity of robust
cooperative strategies.

In contrast to memory capacity, larger games favor defecting strategies over
cooperating strategies, because larger games reduce the marginal cost to a player
of switching from cooperation to defection, and make it harder for even
long-memory players to effectively punish defectors. Thus we find in evolutionary
simulations that only in small games do both long-memory strategies and
cooperation tend to evolve and dominate. It is important to emphasize that these
effects are driven by changes in the volume of robust cooperative strategies. 

A complex balance between behavior, memory, game size and environment can lead to
wide variation in evolutionary outcomes in the presence of social interactions. Understanding this balance is vital if we
are to understand and interpret the role of cooperative behavior in evolution.
Despite the complexity of the problem, and the very general $n$-player memory-$m$
setting we have analyzed, we have arrived at a few simple qualitative predictions,
which may admit to testing not only in the social interactions of natural populations
\cite{Cordero:2012vn} but also through experiments with human players
\cite{Hilbe:2014ab,Rand:2012aa}. Of course, the type of memory discussed here is
only a small part of the story. We have ignored the possibility of other kinds of
memory, which allow players to ``tag'' one another \cite{Adami,Lee:2015aa} after
the completion of a game. We have ignored the role of spatial structure, of
demographic structure, and of dispersal \cite{Rousset:2004aa}. We have failed to
specify the underlying mechanisms by which public-goods and players' decisions are
produced and executed. Accounting for all of these additional factors is an
important challenge as researchers seek to elucidate the emergence of collective
action in evolving populations and beyond.

\clearpage

\begin{flushleft}
{\huge
\textbf{Small games and long memories promote cooperation: Supporting information}
}
\bigskip
\\
Alexander J. Stewart$^{*1,2}$, 
Joshua B. Plotkin$^{1}$
\\
\bigskip
\bigskip
$^1$ Department of Biology, University of Pennsylvania, Philadelphia, PA 19104, USA
\\
$^2$ Current address: Department of Genetics, Environment and Evolution, University College London, London, UK
\\
$^*$ E-mail: alstew@sas.upenn.edu
\\
\end{flushleft}

\vspace{1cm}





\section{Overview of Supporting Information}

In this supplement we detail our analysis of iterated $n$-player games in which players have two
choices in each round and can remember the outcomes of the previous $m$ rounds. 
We identify the strategies that are able to resist
selective invasion by any other strategy in an evolving population of
players.  Such strategies are called ``evolutionary robust'', as defined formally
below.  An iterated  $n$-player game consists of an infinite series of ``rounds''
in each of which each player chooses to either ``cooperate'' ($c$) or ``defect'' ($d$).  A
memory-$m$ strategy stipulates that the probability of cooperation in the
current round depends on the outcomes of the preceding $m$ rounds.  The full space
of memory-$m$ strategies in such an $n$-player game thus has dimension $2^{n\times m}$.  To
identify strategies that are evolutionary robust across such a large space we
first introduce a convenient coordinate transform for the space of memory-$m$
strategies, which generalizes that introduced to study memory-1 strategies in iterated 2-player games \cite{Press:2012fk,Akin,Stewart:2014aa}.  
This coordinate transformation
enables us to identify sets of memory-$m$ strategies that are robust to invasion by any
other strategy in an evolving population. We apply this method to analyse
evolutionary robustness in various $n$-player iterated public goods game.

\subsection{Iterated $n$-player games}

We consider an iterated game with an infinite number of successive rounds between
a player, $X_0$ and her opponents $X_1,X_2. \ . \ .X_{n-1}$. We study games for
which, in each round, each player has two choices, denoted cooperate $(c)$ and
defect $(d)$. The payoffs in a given round to the focal player $X_0$ is given by $R_{c,l-1}$, if
she cooperates along with $l-1$ of her opponents, and it is given by $R_{d,l}$ if
she defects while $l$ of her opponents cooperate.

We will focus on public goods-type games,
for which by definition in each round
\begin{itemize}
\item $R_{d,l}>R_{c,l-1}$ so that, given $l$ players cooperating in total, those who defected 
receive a higher payoff than those who cooperated
\item $R_{c,l}\geq R_{c,l-1}$ and $R_{d,l}\geq R_{d,l-1}$ so that, typically, the more of her opponents cooperate, 
the higher the payoff a cooperative focal player receives.
\end{itemize}

\noindent We will focus in particular on the most typical type of public goods game,
for which $R_{c,l-1}=B\frac{l}{n}-C$ and $R_{d,l}=B\frac{l}{n}$, where $B>C$.

\subsection{Memory-$m$ strategies} 

A memory-$m$ strategy takes account of the outcomes of the preceding $m$ rounds of
play among all players.  As such in any given round there are $n\times m$ plays
taken into account, and the strategy space therefore has dimension $2^{n\times m}$
-- that is, a player's strategy consists of $2^{n\times m}$ probabilities for
cooperation. First we develop notation to describe the probability that a focal
player
will cooperate in a focal round, 
given the plays made by all $n$ players over the preceding $m$ rounds.
We denote the sequence of plays of the $i$th player over
the preceding $m$ rounds $\boldsymbol{\sigma}_i$, which has elements $\sigma^i_k$,
denoting the  play of player $i$, $k$ steps in the past, where $i=0\ldots n-1$ and
$k=1\ldots m$. Thus $\sigma^i_k=c$ if player $i$ cooperated and  $\sigma^i_k=d$ if
she defected $k$ steps in the past. We then write the probability for cooperation for a particular
history of play in its most general form as
$p_{\boldsymbol{\sigma}_0,\boldsymbol{\sigma}_1,\boldsymbol{\sigma}_2,\ .\ .\ .\
,\boldsymbol{\sigma}_{n-1}}\in[0,1]$.

In order to determine the robustness of such strategies, it will be convenient to introduce the operator $\theta$ which returns

\[
	\theta\left(\sigma^i_k\right) = \begin{cases}
		1 & \text{if $\sigma^k_i=c$}\\
		0 & \text{if $\sigma^k_i=d$}
	\end{cases}
\]
\\
where for simplicity we will often write $\theta^i_k$ in place of $\theta\left(\sigma^i_k\right)$ for the play of the $i$th player $k$ steps back in time. The number of times player $i$ cooperated within memory is thus $\sum_{k=1}^m \theta^i_k$ and the number of players who cooperated in the immediately preceding round is $\sum_{i=0}^{n-1}\theta^i_1$.

 
\subsection{Equilibrium payoffs in Iterated Games} The longterm scores received
by $n$ memory-$m$ players in an infinitely iterated game are calculated from the equilibrium
rates of the different plays. This can be determined from the stationary
distribution a Markov chain on $2^{n\times m}$ states, which correspond to the history of plays across the preceding $m$ rounds. 
In order to do this we write the equilibrium rate of a particular history of plays as $v_{\boldsymbol{\sigma}_0,\boldsymbol{\sigma}_1,\boldsymbol{\sigma}_2,\ .\ .\ .\ ,\boldsymbol{\sigma}_{n-1}}$. 

The essential trick we use to analyze equillibrium payoffs in multi-player games among players with
long-memory strategies is to reduce the problem to an equivalent problem involving more
players each using only
memory-1 strategies. The advantage of the memory-1 setting is that it will allow
us to express equillibrium payoffs in the framework of determinants developed by
Press \& Dyson and others \ref{citations here}.
In particular, given a game among $n$ players who memory-$m$ strategies we
construct an equivalent $n\times m$-player game in which players
use only memory-1 strategies. Of these $n\times m$ players, $n$ are ``real"
players, and they each use a
memory-1 strategy that corresponds precisely to a memory-$m$ strategy in the original
long-memory game, as described above. 
In
order to allow the ``real" memory-1 players to effectively react to the entire history of plays across $m$
prior rounds we construct $m-1$ ``shadow'' players for each real player, who
encode the information of earlier rounds. 
At each round,
the shadow player with index $k>1$ deterministically executes the play of its associated real
player, $k$ rounds in the past.  

Given an $n$-player game among memory-$m$ strategies, we encode the equivalent 
$n\times m$-player game among memory-1 strategies by writing $\mathbf{p}^{i,k}$ as the vector of
all $2^{n\times m} $ probabilities for cooperation for the $i$th player, $k$ steps
in the past. If we order the players such that they are indexed from $j=1\ldots
n\times m$ then the index of player $i,k$ is given by $j=i\times m+k$.  In this
labelling system, $\mathbf{p}^{i,1}$ is the strategy of the $i$th real player, and
$\mathbf{v}$ is the corresponding stationary vector of equilibrium rates of play.
Finally we must encode the ``strategy vector'' of the shadow players, which
encode how a player updates her memory each round. This is simple to do. We
write $\mathbf{p}^{i,k}$ as the ``strategy'' vector which updates the memory of
player $i$, $k$ steps in the past (see Figure.~S1 for illustration). This vector
has entry $1$ if $\theta^i_{k-1}=1$ and $0$ if $\theta^i_{k-1}=0$. Thus the real
strategy of player $i$ consists of probabilities $\mathbf{p}^{i,1}\in[0,1]^{n\times m}$,
whereas a shadow strategy, for which $k>1$, consists of deterministic quantities
$\mathbf{p}^{i,k}\in\left\{0,1\right\}^{n\times m}$.
\\
\\
The equilibrium score of player $X_0$ against players
$X_1,X_2. \ . \ . X_{n-1}$ is calculated according to a particular
form of determinant $D$ defined below and written as:
\begin{equation}
S^0_1=\frac{\mathbf{v}\cdot\mathbf{R}^0_1}{\mathbf{v}\cdot\mathbf{I}}=\frac{D\left(\mathbf{p}^{0,1},\mathbf{p}^{0,2}. \ . \ . \mathbf{p}^{0,m},\mathbf{p}^{1,1},\mathbf{p}^{1,2}. \ . \ . \mathbf{p}^{1,m}. \ . \ .\mathbf{p}^{n-1,1},\mathbf{p}^{n-1,2}. \ . \ . \mathbf{p}^{n-1,m},\mathbf{R}^{0,1}\right)}{D\left(\mathbf{p}^{0,1},\mathbf{p}^{0,2}. \ . \ . \mathbf{p}^{0,m},\mathbf{p}^{1,1},\mathbf{p}^{1,2}. \ . \ . \mathbf{p}^{1,m}. \ . \ .\mathbf{p}^{n-1,1},\mathbf{p}^{n-1,2}. \ . \ . \mathbf{p}^{n-1,m},\mathbf{I}\right)}
\end{equation}
\\
Note that in this expression have used the notation of the associated game with $n
\times m$, memory-1 players, $n$ of which correspond to the players in the
original $n$-player, memory-$m$ game. In this equation $\mathbf{I}$ denotes the identity vector of size $2^{n\times m}$, 
for which all elements are 1, 
and $\mathbf{R}^{0,1}$ denotes the payoff vector of
player $X_0$. The payoff to player $0$ in a given round depends only on her own play and the plays 
of the $n$ other players in that round. In general, the payoffs received by player $i$, in the round
that occurred $k$ steps previously is determined from the payoff vector $\mathbf{R}^{i,k}$, which
has $2^{n\times m}$ elements 
$R^{i,k}_{\boldsymbol{\sigma}_0,\boldsymbol{\sigma}_1,\boldsymbol{\sigma}_2,\ .\ .\ .\ ,\boldsymbol{\sigma}_{n-1}}$ 
can be written as
\begin{equation}
R^{i,k}_{\boldsymbol{\sigma}_0,\boldsymbol{\sigma}_1,\boldsymbol{\sigma}_2,\ .\ .\ .\ ,\boldsymbol{\sigma}_{n-1}}=\begin{cases}
		R_{c,\sum_{j=0}^{n-1}\theta^j_k} & \text{if $\theta^i_k=1$}\\
		R_{d,\sum_{j=0}^{n-1}\theta^j_k} & \text{if $\theta^i_k=0$}
	\end{cases}
\end{equation}
\\
Where for the standard public goods game we can write 
$R_{c,\sum_{j=0}^{n-1}\theta^j_k}=\frac{b}{n}\sum_{j=0}^{n-1}\theta^j_k-c\theta^i_k$ 
and $R_{d,\sum_{j=0}^{n-1}\theta^j_k}=\frac{b}{n}\sum_{j=0}^{n-1}\theta^j_k$.
\\
\\
In general, the determinant $D\left(\mathbf{p}^{0,1},\mathbf{p}^{0,2}. \ . \ . \mathbf{p}^{0,m},\mathbf{p}^{1,1},\mathbf{p}^{1,2}. \ . \ . \mathbf{p}^{1,m}. \ . \ .\mathbf{p}^{n-1,1},\mathbf{p}^{n-1,2}. \ . \ . \mathbf{p}^{n-1,m},\mathbf{f}\right)$ arises from a generalization of the results of Press \& Dyson, \cite{Press:2012fk} for two-player games, and of \cite{HimbeMulti,PanMulti} for multiplayer games, and gives the dot product between the stationary vector $\mathbf{v}$ and an
arbitrary vector $\mathbf{f}$ which has elements $f_{\boldsymbol{\sigma}_0,\boldsymbol{\sigma}_1,\boldsymbol{\sigma}_2,\ .\ .\ .\ ,\boldsymbol{\sigma}_{n-1}}$.
In the example of a three player game with memory-1 strategies between players $X_0$, $X_1$ and $X_2$ with strategies $\mathbf{p}$, $\mathbf{q}$ and $\mathbf{r}$, the determinant is given by

\footnotesize\begin{align}
\nonumber D(\mathbf{p_x},\mathbf{q_y},\mathbf{r_z},\mathbf{f})=\\
\det\left[ \begin{array}{cccccccc}
-1+p_{c,c,c}q_{c,c,c}r_{c,c,c} & -1+p_{c,c,c}q_{c,c,c} & -1+p_{c,c,c}r_{c,c,c} & -1+p_{c,c,c} &-1+q_{c,c,c}r_{c,c,c}&-1+q_{c,c,c} & -1+r_{c,c,c} & f_{c,c,c} \\
p_{c,c,d}q_{c,c,d}r_{d,c,c} & -1+p_{c,c,d}q_{c,c,d} & p_{c,c,d}r_{d,c,c} & -1+p_{c,c,d} & q_{c,c,d}r_{d,c,c}&-1+q_{c,c,d} & r_{d,c,c} & f_{c,c,d} \\
p_{c,d,c}q_{d,c,c}r_{c,d,c} & p_{c,d,c}q_{d,c,c} & -1+p_{c,d,c}r_{c,d,c} & -1+p_{c,d,c} &q_{d,c,c}r_{c,d,c}&q_{d,c,c} & -1+r_{c,d,c} & f_{c,d,c} \\
p_{c,d,d}q_{d,c,d}r_{d,d,c} & p_{c,d,d}q_{d,c,d} & p_{c,d,d}r_{d,d,c} & -1+p_{c,d,d} &q_{d,c,d}r_{d,d,c}&q_{d,c,d} & r_{d,d,c} & f_{c,d,d} \\
p_{d,c,c}q_{c,d,c}r_{c,c,d} & p_{d,c,c}q_{c,d,c} & p_{d,c,c}r_{c,c,d} & p_{d,c,c} &-1+q_{c,d,c}r_{c,c,d}&-1+q_{c,d,c} & -1+r_{c,c,d} & f_{d,c,c} \\
p_{d,c,d}q_{c,d,d}r_{d,c,d} & p_{d,c,d}q_{c,d,d} & p_{d,c,d}r_{d,c,d} & p_{d,c,d} &q_{c,d,d}r_{d,c,d}&-1+q_{c,d,d} & r_{d,c,d} & f_{d,c,d} \\
p_{d,d,c}q_{d,d,c}r_{c,d,d} & p_{d,d,c}q_{d,d,c} & p_{d,d,c}r_{c,d,d} & p_{d,d,c} &q_{d,d,c}r_{c,d,d}&q_{d,d,c} & -1+r_{c,d,d} & f_{d,d,c} \\
p_{d,d,d}q_{d,d,d}r_{d,d,d} & p_{c,d,d}q_{c,d,d} & p_{c,d,d}r_{c,d,d} & p_{c,d,d} &q_{c,d,d}r_{c,d,d}&q_{c,d,d} & r_{c,d,d} & f_{d,d,d} \\
 \end{array} \right].
\end{align}\normalsize
 \\
Eq.~1 can be used to calculate the scores received by $n$ memory-1 players in a given game. However, there are certain cases in which the Markov chain
describing the iterated game has multiple absorbing states, and the denominator of Eq.1 goes to zero. The scores in these cases can be calculated by assuming that
players execute their strategy with some small ``error rate'' $\epsilon$ \cite{Fund2}, so that the probability of cooperation is at most $1-\epsilon$
and at least $\epsilon$. Assuming this, and taking the limit $\epsilon\to0$ then gives the player's scores in the cases where multiple absorbing
states exist.

\section{Evolution in a population of players}

We study the evolution of memory-$m$ strategies in a population of $N$ individuals
playing an iterated $n$-player game, with $N\geq n$.  In each generation, all
subsets of $n$ players in the population engage in the iterated game, 
and each player in the
population receives a total score across all the $\binom{N-1}{n-1}$ 
games in which she participates.  We
assume that the population is well-mixed, so that the makeup of different
strategies these games depends upon the frequencies of strategies in the
population.  We focus on evolution under weak-mutation, in which a strategy $X$ is
resident in the population; a single mutant strategy $Y$ arises through
mutation; and $Y$ is 
subsequently either lost or goes to fixation in the population, before another mutant arises.
We always use $X$ to denote the resident, and $Y$ the mutant, strategy.  Under
this weak-mutation assumption there are at most two strategy types present in 
the population at any time.

We use the notation $S^X_a$ to denote the payoff to strategy $X$ in a single iterated
game involving $a$ players of type $Y$ and $n-a$ players of type $X$. We use
the notation $S^Y_a$ to denote the payoff to strategy $Y$ in a single iterated
game involving $a$ players of type $Y$ and $n-a$ players of type $X$.
When the population as a whole contains $b$ players of type $Y$ and $N-b$ players of type
$X$, then, the total score to a player of type $X$, denoted $T^X(b)$, is given by
%
\[
T^X(b)=\frac{(N-n)!(n-1)!}{(N-1)!}\sum_{a=0}^{\min[b,n-1]}\frac{(N-1-b)!}{(N-b-(n-a))!(n-1-a)!}\frac{b!}{a!(b-a)!}S^X_{a}
\] 
where the sum over $a$ denotes the different number of opponents of type $Y$ that $X$ may
face in the $n$-player games she plays in a single generation.
The total score to a mutant $Y$ in such a population, denoted $T^Y(b)$, can be
calculated in the same way.

We model evolution according to the copying process \cite{Traulsen:2006zr}, in which pairs of players are drawn at random from the population, 
and the first player switches her strategy to that of the second player with a
probability that depends on the difference between their total scores. 
Thus a player using a strategy $X$ switches to $Y$ with probability
\[
f_{X\to Y}=\frac{1}{1+\exp\left[s\left(T^X(a)-T^Y(a)\right)\right]}
\]
\\
where $s$ is a parameter denoting the strength of selection. 

The ``strong-selection" regime of this process occurs when $Ns\gg1$.
Under this regime selection is sufficiently strong that an invading mutant
is extremely unlikely to reach high frequency in the populaiton, unless it has a selecive advantage (or is neutral)
against the resident strategy in the population. Thus under strong selection, resident strategies that can
resist invasion by all other mutants are evolutionary robust. Alternatively, the
``weak-selection" limit arises when $Ns\ll1$ in which case even deleterious strategies may reach high
frequency through genetic drift. We focus on the regime of strong selection in
our analysis below.

\subsection{Evolutionary robustness}

The concept of evolutionary robustness\cite{Stewart:2013fk} is similar to the notion of evolutionary stability
\cite{Maynard,Maynard-Smith:1982vn}, but more useful for studying evolution in large strategy spaces, in which an ESS strategy typically does not exist \cite{Stewart:2013fk,Stewart:2014aa}.
In general, a strategy is defined to be evolutionary robust if, when resident in a population, there is no mutant that is favored to spread by 
natural selection when rare \cite{Stewart:2013fk}. 

More precisely, under strong selection a resident strategy $X$ is evolutionary robust iff $T^Y(1) \leq T^X(1)$ for all strategies $Y$.  This condition
for evolutionary robustness under strong selection is identical to that of a Nash equilibrium in the limit $N\to\infty$.


\section{Coordinate Transform}

In two-player games, the work of Press \& Dyson \cite{Press:2012fk} and Akin \cite{Akin} allows us to identify
a coordinate transform for the full space of memory-1 strategies. 
This coordinate transformation permits a simple closed-form expression relating
the scores of two players in a game, which has enabled us to identify all 
evolutionary robust memory-1
strategies, under both strong and weak selection \cite{Stewart:2013fk,Stewart:2014aa}. Here we extend this line of
analysis to
multi-player games with memory-$m$ strategies.  We begin by identifying an analogous coordinate transform for
the $2^{n\times m}$-dimensional space of memory-$m$ strategies in an $n$-player game. 

To define the desired co-ordinate transform we must
identify $2^{n\times m}$ vectors that form a basis in $\mathbb{R}^{n\times m}$ and that allow us to
write down a simple, closed-form relationship between the players' scores in a given game.

These vectors consist firstly of
the $n\times m+1$ vectors $\mathbf{R}^i_k$ for each player's payoff in the $k$th preceding round, along with the identify vector
$\mathbf{I}$, with
entry 1 in all positions. The second set of vectors in the coordinate transform
consists of the $n\times m-1$ vectors denoted $\mathbf{L}^{l}$ , where
$\mathbf{L}^{l}$ is has entry 1 when $l$ players cooperated in the previous $m$ rounds
and entry $0$ otherwise, regardless of the focal player's play in the previous
round. Note that this excludes the case where all players cooperated and the case where no players cooperated, which are accounted for by the identity vector.

The final $2^{n\times m}-2n\times m$ vectors required for the coordinate transform account for the
degeneracy that arises due to the number of ways in which $l$ players can cooperate in the preceding $m$ rounds.  
Given $l$, if the focal player cooperates $l_p$ times over $m$ rounds, and her opponents therefore cooperate $l_o=l-l_p$ times, she will receive the same total payoff over those $m$ rounds in a standard public goods game, regardless of which players cooperated or when. In the most general case a player may nonetheless distinguish between the play of each player (including herself) in each of the preceding $m$ rounds.


We already have $2n\times m$ vectors, as described above.
The simplest way to account for the remaining dimensions (required for players to distinguish between all possible outcomes) is simply to add
\[
2\sum_{l=1}^{n\times m-1}\left(\frac{(n\times m-1)!}{k!(n\times m-l-1)!}-1\right)=2^{n\times m}-2n\times m
\]
\\
vectors, denoted $\mathbf{G}^{l_o,l_p}_{\boldsymbol{\sigma}_0,\boldsymbol{\sigma}_1,\boldsymbol{\sigma}_2,\ .\ .\ .\ ,\boldsymbol{\sigma}_{n-1}}$, which have entry $1$ for a single set of plays $\left(\boldsymbol{\sigma}_0,\boldsymbol{\sigma}_1,\boldsymbol{\sigma}_2,\ .\ .\ .\ ,\boldsymbol{\sigma}_{n-1}\right)$, for which $\sum_{i=1}^{n-1}\sum_{k=1}^{m}\theta^i_k=l_o$ is the total number of times all players have cooperated in the last $m$ rounds and $\sum_{k=1}^{m}\theta^0_k=l_p$ is the number of times the focal player has cooperated in the last $m$ rounds. Note that we have written these vectors in terms of $l_p$ and $l_o$ in order to aid later analysis.

We adopt the convention that we do not add a vector $\mathbf{G}^{l_o,l_p}_{\boldsymbol{\sigma}_0,\boldsymbol{\sigma}_1,\boldsymbol{\sigma}_2,\ .\ .\ .\ ,\boldsymbol{\sigma}_{n-1}}$ for the set of opponent plays which is ordered $cccc. \ . \ .dddd$ across the ordered history of all players, and the set focal player plays ordered either $cccc. \ . \ .dddd$ or $dccc. \ . \ .dddd$. That is, the play for which the first $l_o$ terms of the sum  $\sum_{i=1}^{n-1}\sum_{k=1}^{m}\theta^i_k$ are 1, and either $\theta^1_0=1$ and  $\sum_{k=1}^{l_p}\theta^i_k=l_p$ or else  $\theta^1_0=0$ and  $\sum_{k=1}^{l_p+1}\theta^i_k=l_p$.
\\
\\
In summary, the $2^{n\times m}$ vectors for the coordinate transform consist of
\begin{itemize}
\item $n\times m$ vectors $\mathbf{R}$ for the player's payoffs, and the vector $\mathbf{I}$.
\item $n\times m-1$ vectors $\mathbf{L}^l$ with entry 1 when $l$ players cooperated in the previous round.
\item $2^{n\times m}-2n\times m$ vectors $\mathbf{G}$ with a single entry $1$, to account for the degeneracy which arises when different combinations of opponents cooperate.
\end{itemize}
Although this is a somewhat complex transformation, we shall see the utility of working this way in what follows.
\\
\\
For clarity's sake we can write down this coordinate system explicitly, first for the
case of $n=3$ players with memory-1. The new coordinate system is 
$\left\{\mathbf{R}^0_1,\mathbf{R}^1_1,\mathbf{R}^2_1,\mathbf{I},\mathbf{L}^1,\mathbf{L}^2,\mathbf{G}^{1,1}_{c,d,c},\mathbf{G}^{1,0}_{d,d,c}\right\}$ and we have

\begin{equation}
\det\left[ \begin{array}{cccccccc}
R_{c,2} & R_{c,2} & R_{c,2} & 1 & 0 & 0 & 0 & 0 \\
R_{c,1} & R_{c,1} & R_{d,2} & 1 & 1 & 0 & 0 & 0 \\
R_{c,1} & R_{d,2} & R_{c,1}  &  1 & 1 & 0 & 1 & 0 \\
R_{c,0} & R_{d,1} & R_{d,1} & 1 & 0 & 1 & 0 & 0 \\
R_{d,2} & R_{c,1} & R_{c,1} & 1 & 1 & 0 & 0 & 0 \\
R_{d,1} & R_{c,0} & R_{d,1} & 1 & 0 & 1 & 0 & 0 \\
R_{d,1} & R_{d,1} & R_{c,0} & 1 & 0 & 1 & 0 & 1 \\
R_{d,0} & R_{d,0} & R_{d,0} & 1 & 0 & 0 & 0 & 0 \\
\end{array} \right]=-3(R_{c,2}-R_{d,0})(R_{d,1}-R_{c,0})(R_{d,2}-R_{c,1})
\end{equation}
\\
which is therefore a basis $\mathbb{R}^{8}$. Similarly, for the case of $n=2$ players with with memory-$m$ the new coordinate system is 
$\left\{\mathbf{R}^0_1,\mathbf{R}^1_1,\mathbf{R}^0_2,\mathbf{R}^1_2,\mathbf{I},\mathbf{L}^1,\mathbf{L}^2,\mathbf{L}^3,\mathbf{G}^{1,0}_{dd,cd},\mathbf{G}^{1,0}_{dd,dc},\mathbf{G}^{2,0}_{dd,cc},\mathbf{G}^{1,1}_{dc,dc},\mathbf{G}^{1,1}_{cd,cd},\mathbf{G}^{1,1}_{cd,dc},\mathbf{G}^{1,2}_{cc,dc},\mathbf{G}^{2,1}_{cd,cc}\right\}$ and we have

\begin{equation}
\det\left[ \begin{array}{cccccccccccccccc}
R_{c,2} & R_{c,2} & R_{c,2} & R_{c,2} & 1 & 0 & 0 & 0 & 0 & 0 & 0 & 0 & 0 & 0 & 0 & 0 \\
R_{c,2} & R_{c,1} & R_{c,2} & R_{d,1} & 1 & 0 & 0 & 1 & 0 & 0 & 0 & 0 & 0 & 0 & 0 & 0 \\
R_{c,2} & R_{d,1} & R_{c,2} & R_{c,1} & 1 & 0 & 0 & 1 & 0 & 0 & 0 & 0 & 0 & 0 & 0 & 1 \\
R_{c,2} & R_{d,0} & R_{c,2} & R_{d,0} & 1 & 0 & 1 & 0 & 0 & 0 & 0 & 0 & 1 & 0 & 0 & 0 \\
R_{c,1} & R_{c,2} & R_{d,1} & R_{c,2} & 1 & 0 & 0 & 1 & 0 & 0 & 0 & 0 & 0 & 0 & 1 & 0 \\
R_{c,1} & R_{c,1} & R_{d,1} & R_{d,1} & 1 & 0 & 1 & 0 & 0 & 0 & 0 & 0 & 0 & 0 & 0 & 0 \\
R_{c,1} & R_{d,1} & R_{d,1} & R_{c,1} & 1 & 0 & 1 & 0 & 0 & 0 & 0 & 0 & 0 & 1 & 0 & 0 \\
R_{c,1} & R_{d,0} & R_{d,1} & R_{d,0} & 1 & 1 & 0 & 0 & 0 & 0 & 0 & 0 & 0 & 0 & 0 & 0 \\
R_{d,1} & R_{c,2} & R_{c,1} & R_{c,2} & 1 & 0 & 0 & 1 & 0 & 0 & 0 & 0 & 0 & 0 & 0 & 0 \\
R_{d,1} & R_{c,1} & R_{c,1} & R_{d,1} & 1 & 0 & 1 & 0 & 0 & 0 & 0 & 0 & 0 & 0 & 0 & 0 \\
R_{d,1} & R_{d,1} & R_{c,1} & R_{c,1} & 1 & 0 & 1 & 0 & 0 & 0 & 1 & 0 & 0 & 0 & 0 & 0 \\
R_{d,1} & R_{d,0} & R_{c,1} & R_{d,0} & 1 & 1 & 0 & 0 & 1 & 0 & 0 & 0 & 0 & 0 & 0 & 0 \\
R_{d,0} & R_{c,2} & R_{d,0} & R_{c,2} & 1 & 0 & 1 & 0 & 0 & 0 & 0 & 1 & 0 & 0 & 0 & 0 \\
R_{d,0} & R_{c,1} & R_{d,0} & R_{d,1} & 1 & 1 & 0 & 0 & 0 & 0 & 0 & 0 & 0 & 0 & 0 & 0 \\
R_{d,0} & R_{d,1} & R_{d,0} & R_{c,1} & 1 & 1 & 0 & 0 & 0 & 1 & 0 & 0 & 0 & 0 & 0 & 0 \\
R_{d,0} & R_{d,0} & R_{d,0} & R_{d,0} & 1 & 0 & 0 & 0 & 0 & 0 & 0 & 0 & 0 & 0 & 0 & 0 \\
\end{array} \right]=-4(R_{c,2}-R_{d,0})^2(R_{d,1}-R_{c,1})^2
\end{equation}
\\
which is therefore a basis $\mathbb{R}^{16}$.

\bigskip

Using the results of Press \& Dyson, generalised to multi-player games \cite{Press:2012fk,HimbeMulti,PanMulti}, the strategy of the focal player in this coordinate new system, which for convenience we assign index $i=0$, is given by a vector of the
form:

\begin{equation}
\mathbf{p}^{0,1}-\boldsymbol{\theta}^0_1=\sum_{i=0}^{n-1}\sum_{k=1}^{m}\alpha^i_k\mathbf{R}^i_k+\alpha_n\mathbf{1}+\sum_{l_o=0}^{(n-1)\times m}\sum_{l_p=0}^{m}\left[\lambda_{l_o+l_p}\mathbf{L}^{l_o+l_p}+\sum_{\boldsymbol{\sigma}\in\mathcal{H}_{o,p}}\gamma^{l_o,l_p}_{\boldsymbol{\sigma}}\mathbf{G}^{l_o,l_p}_{\boldsymbol{\sigma}}\right].
\end{equation}
\\
where we define $\mathcal{H}_{o,p}=\left\{(\sigma_0,\ldots,\sigma_n)| \sum^m_{k=1}\theta^0_k=l_p,\sum^{n-1}_{i=1}\sum^{n-1}_{k=1}\theta^i_k=l_o\right\}$ is the set of combinations of plays by $n$ players across the last $m$ rounds such that player 0 (the focal player) cooperated $l_p$ times and her oppoennts cooperated a total $l_o$ times. Note that we set $\lambda=\gamma=0$ when $l_o=l_p=0$ and $l_o+l_p=n\times m$. The vector $\boldsymbol{\theta}^0_1$ has the corresponding elements $\theta^0_1$ for the play of the focal player in the preceding round (i.e 1  if she cooperated and 0 if she defected), and we have written $\sigma=(\sigma_0,\ldots,\sigma_n)$.
\\
\\
From Eq.1 the scores of the players are then related by the expression

\begin{equation}
\sum_{i=0}^{n-1}\sum_{k=1}^{m}\alpha^i_kS^i+\alpha_n\mathbf{1}+\sum_{l_o=0}^{(n-1)\times m}\sum_{l_p=0}^{m}\left[\lambda_{l_o+l_p}v^{l_o,l_p}+\sum_{\boldsymbol{\sigma}\in\mathcal{H}_{o,p}}\gamma^{l_o,l_p}_{\boldsymbol{\sigma}}v^{l_o,l_p}_{\boldsymbol{\sigma}}\right]=0.
\end{equation}
\\
where $S^i$ denotes the equilibrium score of player $i$ in the current game,
$v^{l_o+l_p}$ denotes the rate at which $l_o+l_p$ players cooperate and $v^{l_o,l_p}_{\boldsymbol{\sigma}}$ is the rate at which the focal player cooperates $l_p$ times, along with $l_o$ of
her opponents, with the sequence of plays following the ordering $\boldsymbol{\sigma}=(\boldsymbol{\sigma}_0,\boldsymbol{\sigma}_1,\boldsymbol{\sigma}_2,\ .\ .\ .\ ,\boldsymbol{\sigma}_{n-1})$. 
Note that the equilibrium score of player $i$ is independent of $k$ in Eq.~7. 


We now additionally define the parameters $\chi^0_k=-\alpha^0_k$;
$\phi_{i\times m+k}=\alpha^i_k$;
and $\kappa\left(\sum_{k=1}^{m}\left(\chi^0_k-\sum_{i=1}^{n}\phi_{i\times m+k}\right)\right)=\alpha_n$. In this new parameterization we can re-write the relationship among the players'
scores as:

\begin{equation}
\sum_{k=1}^{m}\left(\sum_{i=1}^{n-1}\phi_{i\times m+k}(S^i-\kappa)-\chi^0_k (S^{0}-\kappa)\right)+\sum_{l_o=0}^{(n-1)\times m}\sum_{l_p=0}^{m}\left[\lambda_{l_o+l_p}v^{l_o,l_p}+\sum_{\boldsymbol{\sigma}\in\mathcal{H}_{o,p}}\gamma^{l_o,l_p}_{\boldsymbol{\sigma}}v^{l_o,l_p}_{\boldsymbol{\sigma}}\right]=0
\end{equation}
\\



Eq.8 gives the most general form for the relationship between player's scores in an $n$-player game with memory-$m$. Henceforth we will restrict our analysis restricted to a focal strategy in which a memory-$m$ player does not distinguish
between her opponents, and does not pay attention to the order of cooperation events. As such we consider a focal player who keeps track of two quantities: (i) the total number of times her opponents cooperated in the last $m$ rounds and (ii) the total number of times she cooperated in the last $m$ rounds.

\subsection{Strategies that track cooperation frequency}

If a focal player tracks only the number of times she cooperated in the last $m$ rounds, and the total number of times her opponent cooperated in the last $m$ rounds, then her memory-$m$ strategy consists of $((n-1)m+1)\times (m+1)$, since her $(n-1)$ opponents can cooperate anywhere between $0$ and $(n-1)m$ times in $m$ rounds, and she can cooperate anywhere between $0$ and $m$ times, to give a strategy consisting of $((n-1)m+1)\times (m+1)$ probabilities for cooperation. We will henceforth explicitly adopt a standard public goods payoff structure, with $R_{c,l}=B\frac{l}{n}-C$ and $R_{d,l}=B\frac{l}{n}$
\\
\\
Eq.~6 encodes a strategy with $2^{n\times m}$ probabilities for cooperation, many of which are redundant in our reduced strategy space. Let the focal player cooperate $l_p$ times and her opponents cooperate $l_o$ times in $m$ rounds. 

Starting from Eq.~8 are now able to make two observations:
\\
\noindent (i) If the focal player does not distinguish between opponents, or the order in which they cooperate, then her
payoff, and the equilibrium rate of play $v^{l_o,l_p}$  are the same for all $((n-1)\times m)!\times m!$ orderings of opponents and plays. 
Summing over all possible orderings of opponents and dividing by $((n-1)\times m)!\times m!$ results in the equilibrium scores being related by

\begin{equation}
\phi\sum_{j=1}^{n-1}\frac{S^j}{n-1}-\sum_{k=1}^{m}\chi^0_k( S^{0}-\kappa)-\kappa\phi+\sum_{l_o=0}^{(n-1)\times m}\sum_{l_p=0}^{m}\left[\lambda_{l_o+l_p}+\gamma^{l_o,l_p}\right]v^{l_o,l_p}=0
\end{equation}
\\
where we have set $\phi=\sum_{i=1}^{n-1}\sum_{k=1}^m\phi_{i\times m+k}$ and $\frac{l_p!(m-l_p)!}{m!}\frac{l_o!((n-1)\times m-l_o)!}{((n-1)\times m)!}\sum_{\boldsymbol{\sigma}\in\mathcal{H}_{o,p}}\gamma^{l_o,l_p}_{\boldsymbol{\sigma}}=\gamma^{l_o,l_p}$, and the $\gamma$ terms result from noting that, when summing 
over all orderings of opponents and events, a given term $\gamma^{l_o,l_p}$ is multiplied by each rate of play $v^{l_o,l_p}_{\boldsymbol{\sigma}}$ a total

\[
\frac{m!}{l_p!(m-l_p)!}\frac{((n-1)\times m)!}{l_o!((n-1)\times m-l_o)!}
\]
\\
\\
\\
\noindent (ii) If we then sum over all $\frac{m!}{l_p!(m-l_p)!}\frac{((n-1)\times m)!}{l_o!((n-1)\times m-l_o)!}$ degenerate probabilities for a given $l_o$ and $l_p$ we then arrive at
\[
-\frac{l_p}{m}+p^{l_o,l_p}=-\frac{l_o}{(n-1)\times m}C\phi+B\frac{l_o+l_p}{n\times m}\left(\phi-\sum_{k=1}^{m}\chi^0_k\right)+\sum_{k=1}^{m}\chi^0_kC\frac{l_p}{m}-\left(\phi-\sum_{k=1}^{m}\chi^0_k\right)\kappa+\lambda_{l_o+l_p}+\gamma^{l_o,l_p}.
\]
\\
as the expression for the probability of cooperation given that the focal player cooperated $l_p$ times and her opponents $l_o$ times in the last $m$ rounds, assuming she only tracks cooperation frequency.

\subsection{Boundary conditions}
If we recall our convention that the equation lacking a $\gamma$ is that which is ordered with $cccc.....dddd$ etc we then have the following boundary conditions

\[
-1+p^{l_o,m}=-C\sum_{i=1}^{l_o}\phi_i+B\frac{l_o+l_p}{n\times m}\left(\sum_{i=1}^{l_o}\phi_i-\sum_{k=1}^{m}\chi^0_k\right)+\sum_{k=1}^{m}\chi^0_kC-\left(\sum_{i=1}^{l_o}\phi_i-\sum_{k=1}^{m}\chi^0_k\right)\kappa+\lambda_{l_o+m}
\]
\\
and

\[
-1+p^{0,l_p}=B\frac{l_p}{n\times m}\left(\sum_{i=1}^{l_o}\phi_i-\sum_{k=1}^{m}\chi^0_k\right)+\sum_{k=1}^{l_p}\chi^0_kC-\left(\sum_{i=1}^{l_o}\phi_i-\sum_{k=1}^{m}\chi^0_k\right)\kappa+\lambda_{l_p}
\]
\\
Similarly the term with $dccc.....dddd$ lacks a $\gamma$ terms so that 

\[
p^{l_o,m-1}=-C\sum_{i=1}^{l_o}\phi_i+B\frac{l_o+l_p}{n\times m}\left(\sum_{i=1}^{l_o}\phi_i-\sum_{k=1}^{m}\chi^0_k\right)+\sum_{k=2}^{m}\chi^0_kC-\left(\sum_{i=1}^{l_o}\phi_i-\sum_{k=1}^{m}\chi^0_k\right)\kappa+\lambda_{l_o+m-1}
\]
\\
and

\[
p^{0,l_p}= B\frac{l_p}{n\times m}\left(\sum_{i=1}^{l_o}\phi_i-\sum_{k=1}^{m}\chi^0_k\right)\phi+\sum_{k=2}^{l_p+1}\chi^0_kC-\left(\sum_{i=1}^{l_o}\phi_i-\sum_{k=1}^{m}\chi^0_k\right)\kappa+\lambda_{l_p}
\]
\\
First, combining the two expressions for $p^{0,l_p}$ gives 

\[
\phi C\chi^0_{l_p+1}=1+\phi C\chi^0_{1}
\]
\\
and since this must hold for all $l_p$ we have $\chi^0_{l_p+1}=\chi^0_{m}$ is constant, and

\begin{equation}
\phi=\frac{1}{C(\chi^0_{m}-\chi^0_{1})}
\end{equation}
\\
Substituting these into the general expression for $p^{0,l_p}$ we find

\[
\gamma^{0,l_p}=0
\]
\\
Second, combining the expressions for $p^{l_o,m}$ and $p^{l_o,m-1}$ gives
\[
\phi \gamma^{l_o,m}=\frac{m-1}{m}+\phi C\frac{m-1}{m}(\chi^0_1-\chi^0_m)+\phi\gamma^{l_o,m-1}
\]
\\
Substituting for $\phi$ we then find 

\[
\gamma^{l_o,m}=\gamma^{l_o,m-1}
\]
\\
Finally, since $\gamma^{(n-1)\times,m}=0$ by definition this also implies

\[
\gamma^{(n-1)\times m,m-1}=0
\]
\\
We therefore have $((n-1)\times m+1)(m+1)-(n\times m+2)$ parameters $\gamma^{l_o,l_p}$, plus $n\times m-1$ parameter $\lambda_{l_o,l_p}$, plus 3 parameters $\chi$, $\phi$ and $\kappa$ to give a total $((n-1)\times m+1)(m+1)$ parameters as required.
We can use these boundary conditions for $\gamma$ to construct the inverse coordinate transform. We arrive at the three simultaneous equations which can be solved for for $\kappa$, $\chi$ and $\phi$:
\begin{eqnarray*}
&\sum_{i=1}^{(n-1)\times m-1}\left(p^{i,m}-p^{i,m-1}\right)-(p^{(n-1)\times m,m-1}-p^{0,m})=(n-1)-C(n-1)\chi-C\phi\\
&p^{0,0}=\kappa(\phi-\chi)\\
&p^{(n-1)\times m,m}=1+\kappa(\phi-\chi)-(B-C)(\phi-\chi)
\end{eqnarray*}
\\
with the remaining terms $\Lambda^{l_o,l_p}$ being determined by these three parameters plus $p^{l_o,l_p}$.
Finally, we set

\begin{equation}
\chi=\chi^{0}_1+(m-1)\chi^0_m
\end{equation}
\\
and
\begin{equation}
\Lambda^{l_o,l_p}=\lambda_{l_o+l_p}+\gamma^{l_o,l_p}
\end{equation}
\\
to define a coordinate system
characterized by a vector of $((n-1)\times m+1)(m+1)$ numbers, \\
$(\kappa, \chi,\phi, \Lambda^{0,0},\ldots,\Lambda^{(n-1)\times m,m})$ 
where we have conditions $\Lambda^{0,0}=\Lambda{(n-1)\times m,m}=0$ and a third linear condition as described above.

\subsection{Strategies and payoffs in a public goods game}

We can now write the relationship between the players' scores when players do not pay attention to the identity of their opponents as
\begin{equation*}
\phi\sum_{i=1}^{n-1}\frac{S^i}{n-1}-\chi
(S^{0}-\kappa)-\phi\kappa+\sum_{l_o=0}^{(n-1)\times m}\sum_{l_p=0}^{m}\Lambda^{l_o,l_p}v^{l_o,l_p}=0.
\end{equation*}
\\
In the case when one player uses strategy $Y$ and the rest use strategy $X$ we
then have the following relationship between scores:
\begin{equation}
\phi S^Y\frac{1}{n-1}+\phi S^X\frac{n-2}{n-1}-\chi
(S^{X}-\kappa)-\phi\kappa+\sum_{l_o=0}^{(n-1)\times m}\sum_{l_p=0}^{m}\Lambda^{l_o,l_p}v^{l_o,l_p}=0.
\end{equation}
\\
The strategy of a focal player in a public goods game can then be written as

\begin{eqnarray}
p^{l_o,l_p}=\frac{l_p}{m}+\kappa(\phi-\chi)+\left(B\frac{l_o+l_p}{n\times m}-C\frac{l_p}{m}\right)\chi-\left(B\frac{l_o+l_p}{n\times m}-C\frac{l_o}{(n-1)\times m}\right)\phi-\Lambda^{l_o,l_p}
\end{eqnarray}
\\
Since a viable strategy must have $0\leq p^{l_o,l_p}\leq 1$ we see by looking at $p^{0,0}$ and $p^{(n-1)\times m,m}$ that
\[
0\leq\kappa\leq B-C
\]
\\
and

\[
\phi>\chi
\]
\\
with additional constraints on the other parameters. This in turn implies that a cooperator, for which $p^{(n-1)\times m,m}=1$ necessitates $\kappa=B-C$ and a defector, for which $p^{0,0}=0$ necessitates $\kappa=0$.


\section{Equilibrium rates of play}
We now derive some inequalities that, in combination with Eq.~11, will allow us to
identify the strategies that are evolutionary robust in $n$-player games. 
In general, we can write the score of a focal player with resident strategy $X$ as

\[
S^X=\sum_{l_o=0}^{(n-1)\times m}\sum_{l_p=0}^{m}\left(B\frac{l_o+l_p}{n\times m}-C\frac{l_p}{m}\right)v^{l_o,l_p}
\]
\\
Similarly the score of an opponent with a strategy $Y$, is given by

\[
S^Y=\sum_{l_o=0}^{(n-1)\times m}\sum_{l_p=0}^{m}\left(B\frac{l_o+l_p}{n\times m}-C\frac{l_p}{m}\right)w^{l_o,l_p}
\]
\\
where $w^{l_o,l_p}$ are the equilibrium rates of play from $Y$'s perspective.
When there is only a single $Y$ mutant in a game, then from $Y$'s perspective, all opponents are identical and use strategy $X$. In this situation we can write

\begin{equation}
v^{l_o,l_p}=\sum_{l_p'=\max\left[0,l_o+l_p-(n-1)\times m\right]}^{\min\left[m,l_o\right]}w^{l_o+l_p-l_p',l_p'}\frac{((n-1)\times m-l_o-l_p+l_p')!}{(m-l_p)!((n-2)\times m-l_o+l_p')!}\frac{(l_o+l_p-l_p')!}{l_p!(l_o-l_p')!}\frac{m!((n-2)\times m)!}{((n-1)\times m)!}
\end{equation}
\\
where we assume $w^{l_o+l_p-l_p',l_p'}=0$ for the unphysical case $l_p'>l_o$.
This allows us to write the score of $X$ in terms of $w^{l_o,l_p}$. We will now use these results to explore two special cases of interest: (i) The effect of increasing the size of the game $n$ with fixed memory, and (ii) the effect of increasing memory size $m$ with fixed game size.

\section{Bounds on players' scores}

We can now use Eq.~14 to find upper and lower bounds on the difference and the sum of players' scores, in the case that the game contains a single player using a strategy $Y$ and $n-1$ players using a strategy $X$.

\[
S^X=\sum_{l_o=0}^{(n-1)\times m}\sum_{l_p=0}^{m}\left(B\frac{l_o+l_p}{n\times m}-C\frac{l_o}{(n-1)\times m}\right)w^{l_o,l_p}
\]
\\
We can now write the difference between the scores of $X$ and $Y$ scores as
\[
S^X-S^Y=\sum_{l_o=0}^{(n-1)\times m}\sum_{l_p=0}^{m}C\frac{l_p(n-1)-l_o}{(n-1)\times m}w^{l_o,l_p}
\]
\\
which enables us to identify upper and lower bounds on the difference between two players' scores, namely

\begin{equation}
S^X-S^Y\geq-\sum_{l_o=0}^{(n-1)\times m}\sum_{l_p=0}^{m}C\frac{l_o}{(n-1)\times m}w^{l_o,l_p}
\end{equation}
\\
which becomes an equality when $Y$ always defects at equilibrium
and

\begin{equation}
S^X-S^Y\leq\sum_{l_o=0}^{(n-1)\times m}\sum_{l_p=0}^{m}C\frac{(n-1)\times m-l_o}{(n-1)\times m}w^{l_o,l_p}
\end{equation}
\\
which becomes an equality when $Y$ never defects at equilibrium.
We can similarly write, for the sum of the player's scores,
\[
S^X+S^Y\frac{1}{n-1}=\sum_{l_o=0}^{(n-1)\times m}\sum_{l_p=0}^{m}(B-C)\frac{l_o+l_p}{(n-1)\times m}w^{l_o,l_p}.
\]
\\
This gives an upper bound on the sum 

\begin{equation}
S^X+S^Y\frac{1}{n-1}\leq\frac{n}{n-1}(B-C)-\sum_{l_o=0}^{(n-1)\times m}\sum_{l_p=0}^{m}(B-C)\frac{n\times m-l_o-l_p}{(n-1)\times m}w^{l_o,l_p}.
\end{equation}
\\
which becomes an equality when $w^{0,0}=0$ at equilibrium (i.e it is never the case that all players defect). 
Finally we have a lower bound on the sum

\begin{equation}
S^X_1+S^Y_1\frac{1}{n-1}\geq\sum_{l_o=0}^{(n-1)\times m}\sum_{l_p=0}^{m}(B-C)\frac{l_o+l_p}{(n-1)\times m}w^{l_o,l_p}.
\end{equation}
\\
which becomes an equality when $w^{(n-1)\times m,m}=0$, (i.e it is never the case that all players cooperate at equilibrium).

It is also convenient to rewrite Eq.~12 for the relationship between two player's scores in terms of $w$ to give

\small
\begin{align}
\nonumber S^Y\frac{1}{n-1}+S^X\frac{n-2}{n-1}-(\chi^0_1+(m-1)\chi^0_m) (S^{X}-\kappa)-\kappa+\\
\sum_{l_o=0}^{(n-1)\times m}\sum_{l_p=0}^{m}\left(\sum_{k=\max\left[0,l_o+l_p-(n-1)\times m\right]}^{\min\left[m,l_o\right]}\Lambda^{l_o+l_p-k,k}\frac{((n-1)\times m-l_o)!}{(m-k)!((n-2)\times m-l_o+k)!}\frac{l_o!}{k!(l_o-k)!}\frac{m!((n-2)\times m)!}{((n-1)\times m)!}\right)w^{l_o,l_p}=0.
\end{align}
\normalsize
\\

We can now use Eqs.~16-20 to identify the strategies that are evolutionary robust,
under strong selection, in multi-player games.

\subsection{Robust strategies}

We focus here on the prospects for cooperation in iterated games. In particular, we identify strategies which, when used by all players in a game, ensure that all players cooperate. This is achieved quite simply by setting $p^{n-1,1}=1$ so that if all players cooperated in the preceding round, all players assuredly cooperate in the following round. We call these strategies the cooperators and we calculate the robustness of these strategies by determining the proportion of cooperators that can resist invasion by all other strategies. We contrast this to the defectors: strategies which have $p^{0,0}=0$, such that if al players defected in the preceding round, all players assuredly defect in the following round. The importance of these two strategy classes in two-player public goods games has been established already \cite{Stewart:2014aa}, making it natural to generalise their study to games with multiple players and long memory. 

To determine whether a strategy is robust we use the condition given previously for an evolving population of $N$ players in a multiplayer game. Given
a resident strategy $X$ in a population, selection acts against a new mutant $Y$ provided $T^X(1)>T^Y(1)$, as described above. This can be written explicitly in terms of players' scores as

\begin{equation}
\frac{N-n}{N-1}S^X_{0}+\frac{n-1}{N-1}S^X_{1}>S^Y_{1}
\end{equation}
\\
where $S^X_0$ is the score received by $X$ with no mutants in the game, $S^X_1$ is the score received by $X$ with one mutant player $Y$ in the game and $S^Y_1$ is the score received by $Y$ with no other mutants in the game.

\subsection{Robust cooperating strategies under strong selection}

We first identify the cooperating strategies that are robust under strong
selection. As defined above, a
cooperating strategy $X$ is such that, if all players use the strategy, all
players cooperate every turn at equilibrium. Such strategies must have
$\kappa=B-C$.

A mutant strategy $Y$ can selectively invade a cooperating strategy under strong selection iff

\begin{equation*}
S^Y_{1}-S^X_1>\frac{N-n}{N-1}(B-C-S^X_{1})
\end{equation*}
\\
Thhe longterm payoffs must additionally satisfy Eq.~17-21. We can therefore identify strategies $X$ which cannot be selectively invaded by any mutant $Y$. For simplicity we write

\begin{equation}
\hat{\Lambda}^{l_o,l_p}=\sum_{k=\max\left[0,l_o+l_p-(n-1)\times m\right]}^{\min\left[m,l_o\right]}\Lambda^{l_o+l_p-k,k}\frac{((n-1)\times m-l_o)!}{(m-k)!((n-2)\times m-l_o+k)!}\frac{l_o!}{k!(l_o-k)!}\frac{m!((n-2)\times m)!}{((n-1)\times m)!}.
\end{equation}
\\
\\
\\
\textbf{Case I: Robustness when $\chi\leq\frac{N(n-2)+1}{(N-1)(n-1)}\phi$}
\\
\\
Using Eq.~19 we can write the condition for invasion by a mutant strategy $Y$ as

\begin{eqnarray}
\left(\frac{N(n-2)+1}{(N-1)(n-1)}\phi-\chi\right)(S^X_{1}-(B-C))<-\sum_{l_o=0}^{(n-1)\times m}\sum_{l_p=0}^{m}\hat{\Lambda}^{l_o,l_p}w^{l_o,l_p}
 \end{eqnarray}
\\
combining Eq.~15 with Eq. 19 then gives

\begin{align}
\frac{N-n}{N-1}\sum_{l_o=0}^{(n-1)\times m}\sum_{l_p=0}^{m}\hat{\Lambda}^{l_o,l_p}w^{l_o,l_p}
<C\left(\frac{N(n-2)+1}{(N-1)(n-1)}\phi-\chi\right)\sum_{l_o=0}^{(n-1)\times m}\sum_{l_p=0}^{m}
C\frac{l_o}{(n-1)\times m}w^{l_o,l_p}
\end{align}
\\
as a necessary condition for invasion.
\\
\textbf{Case II: Robustness when $\chi>\frac{N(n-2)+1}{(N-1)(n-1)}\phi$}
\\
\\
Combining Eq.~17 and Eq. 19 we can also write

\begin{align}
\frac{N-n}{N-1}\sum_{l_o=0}^{(n-1)\times m}\sum_{l_p=0}^{m}\hat{\Lambda}^{l_o,l_p}w^{l_o,l_p}
<(B-C)\left(\frac{N(n-2)+1}{(N-1)(n-1)}\phi-\chi \right)\sum_{l_o=0}^{(n-1)\times m}\sum_{l_p=0}^{m}\frac{n\times m-l_o-l_p}{(n-1)\times m}w^{l_o,l_p}
\end{align}
\\
as a necessary condition for robustness.

Thus, in summary, the set of robust cooperating strategies in an $n$-player game under
strong selection with memory-$m$, which we denote $\mathcal{C}^{n,m}_{s}$, is given by: 

\begin{align}
\nonumber  &\mathcal{C}^{n,m}_{s}=\Bigg\{(\chi,\phi,\kappa,
\Lambda^{0,1},\ldots\Lambda^{(n-1)\times m,m-1})\bigg|\kappa=B-C,\\
\nonumber &\frac{N-n}{N-1}\sum_{l_o=0}^{(n-1)\times m}\sum_{l_p=0}^{m}\hat{\Lambda}^{l_o,l_p}w^{l_o,l_p}\geq 
\nonumber  C\left(\phi\frac{N(n-2)+1}{(N-1)(n-1)}-\chi\right)
\sum_{l_o=0}^{(n-1)\times m}\sum_{l_p=0}^{m}\frac{l_o+l_p}{(n-1)\times m}w^{l_o,l_p},\\
\nonumber &\frac{N-n}{N-1}\sum_{l_o=0}^{(n-1)\times m}\sum_{l_p=0}^{m}\hat{\Lambda}^{l_o,l_p}w^{l_o,l_p}\geq 
\nonumber (B-C)\left(\phi\frac{N(n-2)+1}{(N-1)(n-1)}-\chi\right)
\sum_{l_o=0}^{(n-1)\times m}\sum_{l_p=0}^{m}\frac{n\times m-l_o-l_p}{(n-1)\times m}w^{l_o,l_p}\Bigg\}\\
\end{align}
\normalsize
\subsection{Robust defecting strategies under strong selection}

We now identify the defecting strategies that are robust under strong
selection.  As defined above, a defecting strategy $X$ is one such that, if
all players adopt the
strategy, all players defect every turn at equilibrium. Such strategies must have
$\kappa=0$. 

A mutant strategy $Y$ can selectively invade a defecting strategy under strong selection iff

\begin{equation*}
S^X_1-S^Y_{1}<\frac{N-n}{N-1}S^X_{1}
\end{equation*}
\\
Using Eq.~19 this can be re-written as

\begin{align}
S^X_1\left(\frac{N(n-2)+1}{(N-1)(n-1)}\phi-\chi \right)<-\frac{N-n}{N-1}\sum_{l_o=0}^{(n-1)\times m}\sum_{l_p=0}^{m}\hat{\Lambda}^{l_o,l_p}w^{l_o,l_p}
\end{align}
\\
Following the same procedure as for the cooperators above, we find that the set of robust defecting strategies in an $n$-player game under
strong selection with memory-$m$, which we denote $\mathcal{D}^{n,1}_{s}$, is given by 

\begin{align}
&\nonumber  \mathcal{D}^{n,m}_{s}=\Bigg\{(\chi,\phi,\kappa,
\Lambda^{0,1},\ldots\Lambda^{(n-1)\times m,m-1})\bigg|\kappa=0,\\
&\nonumber \frac{N-n}{N-1}\sum_{l_o=0}^{(n-1)\times m}\sum_{l_p=0}^{m}\hat{\Lambda}^{l_o,l_p}w^{l_o,l_p}\geq
\nonumber-(B-C)\left(\phi\frac{N(n-2)+1}{(N-1)(n-1)}-\chi \right)\sum_{l_o=0}^{(n-1)\times m}\sum_{l_p=0}^{m}\frac{l_o+l_p}{(n-1)\times m}w^{l_o,l_p},\\
&\nonumber \frac{N-n}{N-1}\sum_{l_o=0}^{(n-1)\times m}\sum_{l_p=0}^{m}\hat{\Lambda}^{l_o,l_p}w^{l_o,l_p}\geq
\nonumber -C\left(\phi\frac{N(n-2)+1}{(N-1)(n-1)}-\chi\right)\sum_{l_o=0}^{(n-1)\times m}\sum_{l_p=0}^{m}\frac{n\times m-l_o-l_p}{(n-1)\times m}w^{l_o,l_p}\Bigg\}\\
\end{align}
\\
Notice that in the special case $n=N$, in which all members of a population play the same public goods game together, the conditions for robustness are independent of $\Lambda^{l_o,l_p}$.

\subsection{Calculating robust volumes}

We have now derived necessary and sufficient conditions for cooperators and
defectors to be robust in $n$-player public goods games. However, in contrast to
the case of two-player games these conditions depend explicitly on the equilibrium
play $w^{l_o,l_p}$ of an invading mutant. We can nonetheless easily construct
strategies that are assuredly robust, by using the fact that $w^{l_o,l_p}\leq 1$
for all possible mutants. Similarly we can construct strategies that are assuredly
invadable. However this leaves a large subset of strategies whose robustness
depends on the actual values of $w^{l_o,l_p}$. Nonetheless we can still determine
their robustness by using the fact that the bounds on players scores render the
conditions Eqs.~26 and 27 most stringent when a mutant plays such that (1) he
never cooperates at equilibrium (Eq.~16), (2) he plays so that $w^{0,0}=0$ at
equilibrium, i.e so that all players do not defect simultaneously (Eq.~18), (3) he
always cooperates at equilibrium (Eq.~17) or (4) he plays such that all players do
not cooperate simultaneously (Eq.~19). This leaves us with four possible trigger
strategies to test in order to determine the robustness of a cooperator or
defector strategy (where the relevant trigger strategy depends on the values of
$\phi$ and $\chi$ and whether the resident is a cooperator or a defector).  

Figure 3 of the main text verifies the use of these four trigger strategies by
comparing the analytically predicted volumes of robust strategies to those
estimated by Monte Carlo against a large number of randonly chosent mutant
invaders. This type of Monte Carlo verification is also shown in Figure S1 for the
effect of population size $N$ on the volume of robust strategies. As discussed in
the main text, larger populations leader to larger volumes of robust cooperators
and smaller volumes of robust defectors.

\subsection{The impact of memory on robustness}

As discussed in the main text, the impact of memory on robustness arises because it increases the capacity for contingent punishment, as expressed through the parameters $\Lambda^{l_o,l_p}$. The way in which this occurs is most clearly understood by looking at the expectation $\left<\Lambda^{l_o,l_p}\right>$ for a randomly drawn strategy. For a randomly drawn cooperating or defecting strategy the expectations $\left<\chi\right>$ and $\left<\phi\right>$ are related according to
\begin{eqnarray*}
&(n-1)\left<\chi\right>=\frac{(n-1)}{C}-\left<\phi\right>\\
&1/2=(B-C)(\left<\phi\right>-\left<\chi\right>)\\
\end{eqnarray*}
\\
which gives

\begin{eqnarray*}
&\left<\chi\right>=\frac{n-1}{Cn}-\frac{1}{2(B-C)n}\\
&\left<\phi\right>=\frac{n-1}{Cn}+\frac{n-1}{2(B-C)n}\\
\end{eqnarray*}
\\
The expectation $\left<\Lambda^{l_o,l_p}\right>$ for a randomly drawn cooperator is then
\[
\left<\Lambda^{l_o,l_p}\right>=\frac{1}{2}\frac{l_o+l_p}{n\times m}
\]
\\
which gives an average across all $l_o,l_p$ of 

\[
\left<\Lambda\right>=\frac{1}{4}
\]
\\
Similarly, the expectation $\left<\Lambda^{l_o,l_p}\right>$ for a randomly drawn defector is and the average $\Lambda$ is

\[
\left<\Lambda^{l_o,l_p}\right>=-\frac{1}{2}\frac{n\times m-(l_o+l_p)}{n\times m}
\]
\\
which gives an average across all $l_o,l_p$ of 

\[
\left<\Lambda\right>=-\frac{1}{4}
\]
\\
If we now use Eq.~22 to determine the average $\left<\hat{\Lambda}^{l_o,l_p}\right>$ for a cooperators and defectors faced with a mutant who cooperated $l_p$ times within their memory, we recover Fig S2. We see that a randomly drawn cooperator tends to be more succesful at punishing a given mutant, while a randomly drawn defector tends to become less successful, as memory increases.

\subsection{Invasability and cost of memory}

Our evolutionary simulations, Figure 4, show that in addition to increasing the
overall robustness of cooperation, memory capacity $m$ tends to increase in small games. To
understand this we must look at the average fixation probability of mutations that
increase memory by 1, versus those that decrease memory by 1.  This is shown in Figure
S2c. We see that mutations that increase memory capacity are more likely to fix than
mutations that decrease memory capacity, regardless of the current resident memory
capacity. Thus longer memories will tend to evolve on average. If we
introduce a cost for memory, so that a player's overall payoff is reduced by a
factor $mC_m$, we see (Figure S2d) that mutations that increase memory eventually
become worse invaders than mutations that decrease memory. In such cases an
intermediate memory length evolves. Thus the evolution of memory depends on the
costs associated with longer memories, as well as the size of the game being
played. As we see in Figure S2a, shorter memories evolve, and much more
slowly, when memory comes at a cost. Correspondingly (Figure S2b), the effect of
evolving longer memories has a much weaker effect on the evolution of
cooperation, although the general trend of reduced defection and increased
cooperation is maintained.

\subsection{Robustness and the dimension of strategy space}

As discussed in the main text and shown in Figure 4, as memory increases the
overall frequency of robust cooperators and defectors that evolve tends to
decline. This reflects the fact that the absolute volume of robust strategies
tends to decline as the dimension of strategy space increases - the probability of
randomly drawing a strategy from the $n$-dimensional unit cube, that also lies
within a robust volume volume with sides of fixed length, declines as a power of
$1/n$. This decline in the robust volumes of strategies with the dimension of
strategy space (both game size $n$ and memory length $m$) is shown in Figure S4.

\clearpage

\section{Supplementary Figures}

\begin{figure*}[h!] \centering \includegraphics[scale=0.5]{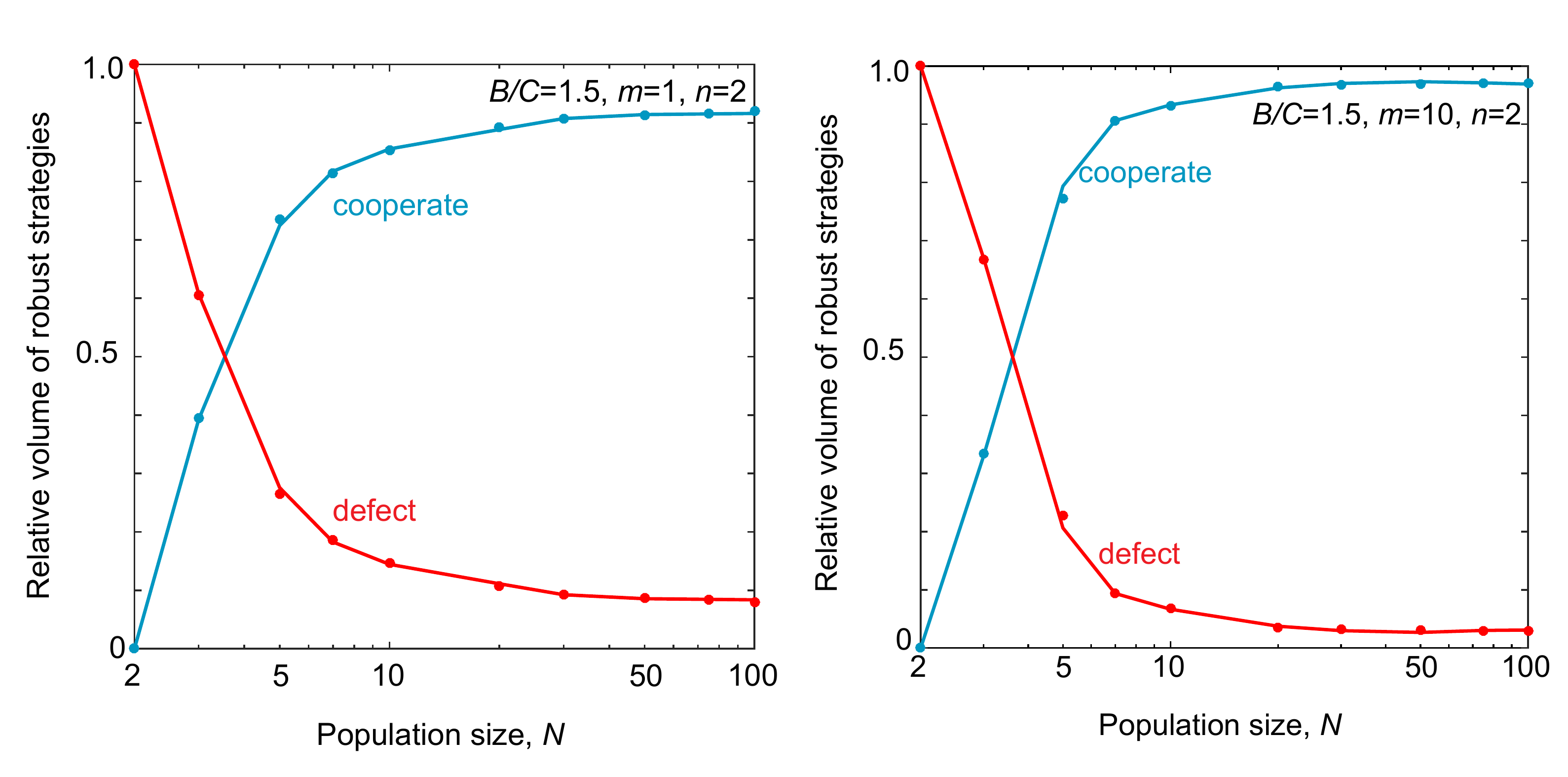}
\caption*{Figure S1: The impact of population size on cooperation. We 
calculated the relative volumes of robust cooperation -- that is, the absolute
volume of robust cooperative strategies divided by the total volume of robust
cooperators and defectors -- and compared this to the relative volume of
defectors (solid lines) using Eqs.~2-3. We also
verified these analytic results by randomly drawing $10^6$ strategies and
determining their
success at resisting invasion from $10^5$ random mutants (points). We calculated player's
payoffs by simulating $2\times 10^3$ rounds of a public-goods game. We then
plotted the relative volumes of robust cooperators and robust defectors as a
function of populations size $N$ with fixed game size $n=2$ and memory length $m=1$, (left) and $m=10$ (right).
In both cases the effect of increasing population size is to increase the relative volume of cooperators and decrease that of defectors.
In all calculations and simulations we used cost $C=1$ and benefit $B$
as indicated in the figure
} 
\end{figure*}

\begin{figure*}[h!] \centering \includegraphics[scale=0.5]{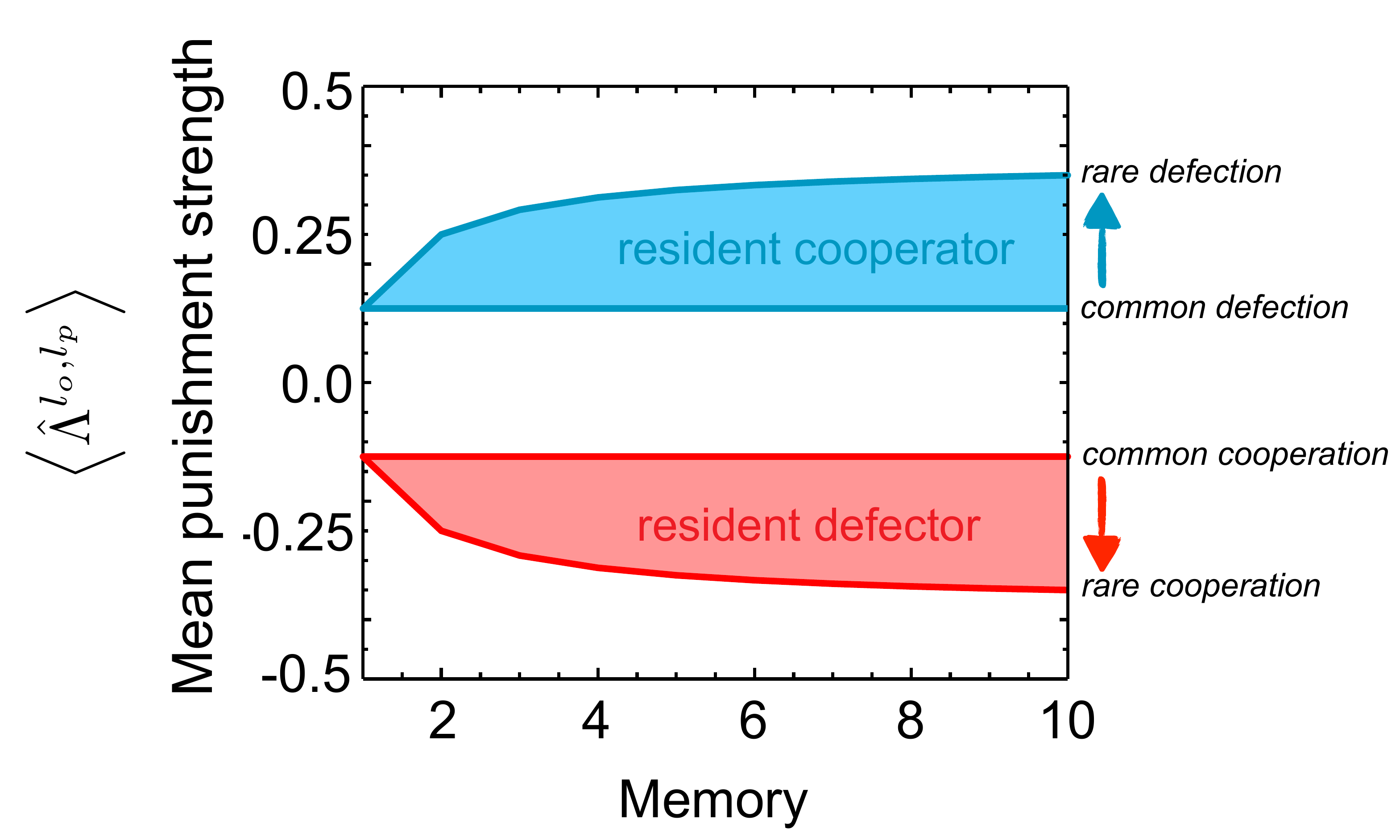}
\caption*{Figure S2: Effectiveness of contingent punishment. We calculated $\frac{1}{(n-1)\times m}\sum_{lo}\left<\hat{\Lambda}^{l_o,l_p}\right>$ for the average punishment of a mutant who defected $l_p$ times within the memory of the resident strategy, for both cooperators (blue) and defectors (red). As memory becomes longer, the average punishment increases for cooperators, making strategies more likely to be robust, and decreases for defectors, making strategies less likely to be robust.
} 
\end{figure*}

\begin{figure*}[h!] \centering \includegraphics[scale=0.5]{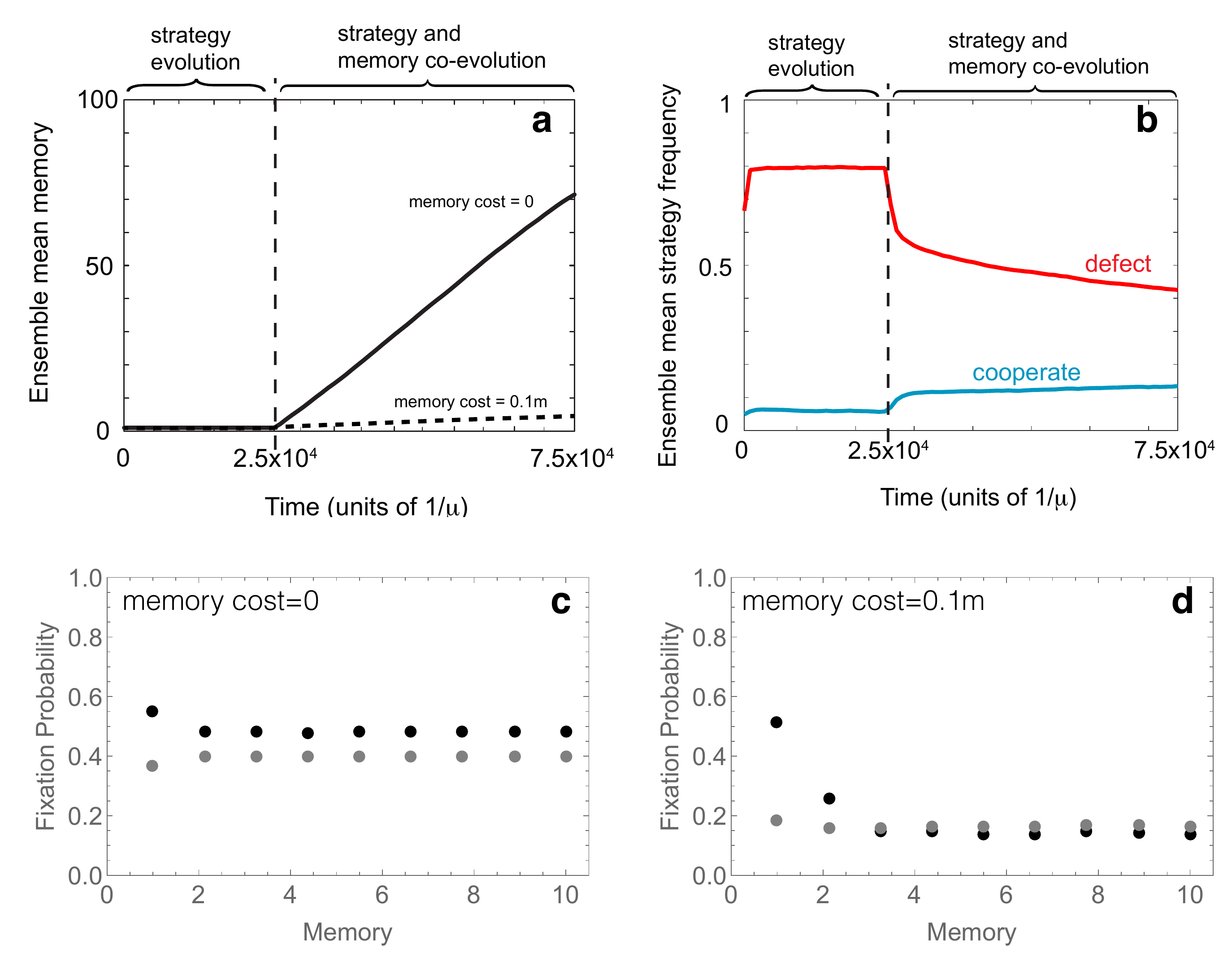}
\caption*{Figure S3: Invasibility of memory. We simulated co-evolution of memory and strategies as 
described in Figure 4 of the main text, with an additional cost to having
memory which reduces a player's payoff by $c_m\times m$. We see that (a) much
shorter memories evolve for $c_m=0.1$ compared to $c_m=0$ and (b) a
correspondingly smaller amount of cooperation evolves. In order to understand why
longer memory strategies evolve in small games we looked at the average fixation
probability of mutations that increase or decrease memory, when played against a
randomly drawn resident strategy. We drew $10^6$ resident strategies for each
memory length $m \in \{1,2,3,/ldots, 10\}$ and for each drew $10^5$ mutants that increase memory length
by 1 and $10^5$ mutants that decrease memory length by 1. We assumed that a
mutation that increased memory length by 1 did not change the probability
$p^{l_o,l_p}$ of the player's strategy. Where mutations increased memory length, we
randomly drew probabilities $p^{(n-1)(m+1),l_p}$ and $p^{l_o,m+1}$. (c) Plotted
are the average fixation probabilities for mutations that increase (black dots) or decrease
(gray dots) memory by 1. Each point shows the probability of in versus out transition
for the state $k$ (i.e mutations that result in increase in memory from $k$ to
$k+1$ and mutations that result in decrease in memory from $k+1$ to $k$). When
there is no cost to memory, mutations that increase memory length are always
better invaders, for games of size $n=2$. (d) When there is a cost to memory,
mutations that decrease memory length do relatively better, and mutations that
increase memory length do relatively worse. As a result we expect to see
intermediate memory lengths evovle in the presence of costs.
} 
\end{figure*}

\begin{figure*}[h!] \centering \includegraphics[scale=0.5]{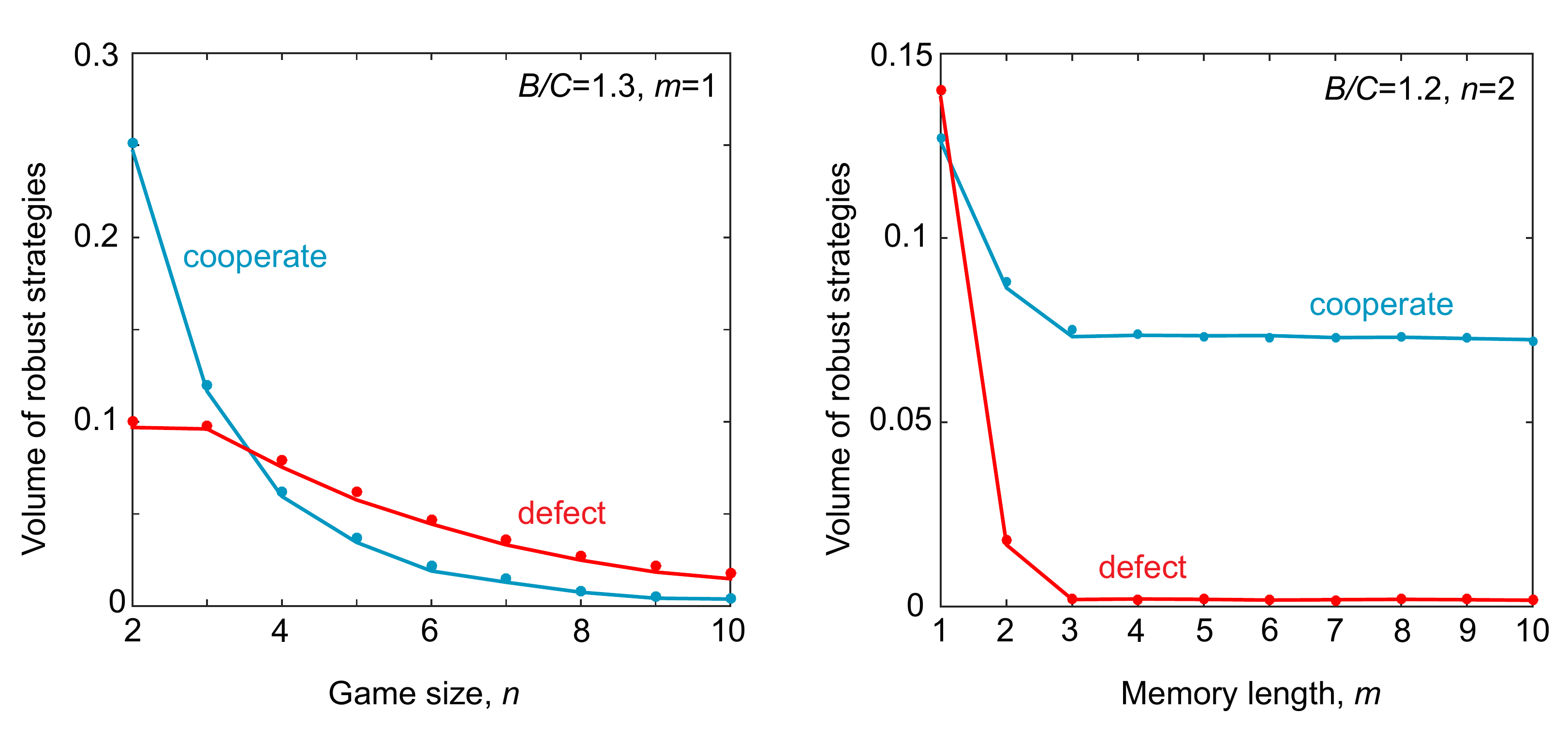}
\caption*{Figure S4: Absolute volumes of robust strategies. Here we show the same plot as in Figure 3 of the main text, using absolute rather than relative volumes. As is clear, the absolute volumes of both cooperators and defectors tends to decline as the dimension of strategy space increases. However this occurs at different rates for the different strategy types depending on whether game size (left) or memory (right) is increasing.} 
\end{figure*}

\clearpage


\end{document}